\documentclass[iop,numberedappendix]{emulateapj}

\slugcomment{Accepted for publication in The Astrophysical Journal} 

\shorttitle{Neutronization in Type Ia Supernovae}
\shortauthors{H. Mart\'{i}nez-Rodr\'{i}guez, A. L. Piro, J. Schwab, and C. Badenes}

\usepackage{graphicx}	
\usepackage{amsmath}	
\usepackage{amssymb}	
\usepackage[dvipsnames]{xcolor}
\usepackage{hyperref} 
\hypersetup{
	colorlinks   = true, 
	urlcolor     = blue, 
	linkcolor    = blue, 
	citecolor   = blue 
}

\usepackage{float}

\newcommand{\be}{\begin{eqnarray}}
\newcommand{\ee}{\end{eqnarray}}

\newcommand{\cm}{\,{\rm cm}}
\newcommand{\g}{\,{\rm g}}

\newcommand{\me}{\ensuremath{m_\mathrm{e}}} 
\newcommand{\mue}{\ensuremath{\mu_\mathrm{e}}} 

\begin{document}


\title{Neutronization During Carbon Simmering In Type Ia Supernova Progenitors}


\author{H\'{e}ctor Mart\'{i}nez-Rodr\'{i}guez\altaffilmark{1}, Anthony L. Piro\altaffilmark{2}, Josiah Schwab\altaffilmark{3,4}, and Carles Badenes\altaffilmark{1}}



\altaffiltext{1}{Department of Physics and Astronomy and Pittsburgh Particle Physics, Astrophysics and Cosmology Center (PITT PACC), University of Pittsburgh, 3941 O'Hara Street, Pittsburgh, PA 15260, USA, \email{hector.mr@pitt.edu}}
\altaffiltext{2}{Carnegie Observatories, 813 Santa Barbara Street, Pasadena, CA 91101, USA}
\altaffiltext{3}{Department of Physics, University of California, Berkeley, CA 94720, USA}
\altaffiltext{4}{Department of Astronomy and Theoretical Astrophysics Center, University of California, Berkeley, CA 94720, USA}


\begin{abstract}
When a Type Ia supernova (SN Ia) progenitor first ignites carbon in its core, it undergoes ${\sim} \,10^{3}-10^{4} \,$years of convective burning prior to the onset of thermonuclear runaway. This carbon simmering phase is important for setting the thermal profile and composition of the white dwarf. Using the \texttt{MESA} stellar evolution code, we follow this convective burning and examine the production of neutron-rich isotopes. The neutron content of the SN fuel has important consequences for the ensuing nucleosynthesis, and in particular, for the production of secondary Fe-peak nuclei like Mn and stable Ni. These elements have been observed in the X-ray spectra of SN remnants like Tycho, Kepler, and 3C 397, and their yields can provide valuable insights into the physics of SNe Ia and the properties of their progenitors. We find that weak reactions during simmering can at most generate a neutron excess of ${\approx} \, 3 \times 10^{-4}$. This is ${\approx} \, 70 \%$ lower than that found in previous  studies that do not take the full density and temperature profile of the simmering region into account. Our results imply that the progenitor metallicity is the main contributor to the neutron excess in SN Ia fuel for $Z \gtrsim 1/3 \, Z_{\odot}$. Alternatively, at lower metallicities, this neutron excess provides a floor that should be present in any centrally-ignited SN~Ia scenario.
\end{abstract}


\keywords{convection -- nuclear reactions, nucleosynthesis, abundances -- methods: numerical -- stars: evolution -- supernovae: general -- white dwarfs}



\section{Introduction} \label{Introduction}

Type Ia supernovae (SNe Ia) are the thermonuclear explosions of white dwarf (WD) stars \citep{Ma14}. They play a key role in galactic chemical enrichment through Fe-peak elements \citep{Iw99}, as cosmological probes to investigate dark energy \citep{Ri98,Pe99} and constrain $\Lambda$CDM parameters \citep{Be14,Re14}, and as sites of cosmic ray acceleration along with other SN types \citep{Ma14}. However, the exact nature of their progenitor systems remains mysterious. While it is clear that the exploding star must be a C/O WD in a binary system \citep{Bl12}, decades of intensive observational and theoretical work have failed to establish whether the binary companion is a non-degenerate star (the so-called single degenerate, or SD, scenario), another WD (double degenerate, DD -- see \citealt{Wa12,Ma14} for recent reviews), or some combination of scenarios. Both cases result in the explosion of a relatively massive WD after one or potentially many more mass accretion episodes, but there are key differences between them. In the SD scenario, the accretion happens over relatively long timescales \citep[$\sim 10^{6} \, $yr,][]{Ha96,Ha04} until the mass of the WD gets close to the Chandrasekhar limit ($M_{\rm{Ch}}$ $= 1.45(2Y_{e})^{2} \approx 1.39 \, M_{\odot}$, where $Y_{e}$ is the mean number of electrons per baryon, \citealt {No84,Th86,Ha96,Ha04,Si10}). In the DD scenario, the explosion is the result of the violent interaction or merging of two WDs on a dynamical timescale \citep{Ib84}, and the mass of the exploding object is not expected to be directly tied to $M_{\rm{Ch}}$ \citep{Si10,vaK10}. Attempts to discriminate between SD and DD systems based on these differences have had varying degrees of success. On the one hand, it is known that WDs in the Milky Way merge at a rate comparable to SN Ia explosions \citep{Ba12}, and statistical studies of the ejecta and $^{56}$Ni mass distribution of SN Ia indicate that a significant fraction of them are not near-Chandrasekhar events \citep{Pi14,Scalzo14}. On the other hand, the large amount of neutron-rich material found in solar abundances \citep{Se13} and in some supernova remnants (SNRs) believed to be of Type Ia origin \citep{Ya15} seems to require burning at high densities, which indicates that at least \textit{a non-negligible fraction} of SNe Ia explode close to $M_{\rm{Ch}}$.

Here we focus on the role that these neutron-rich isotopes play as probes of SN Ia explosion physics and progenitor evolution channels. In particular, we explore the effect of carbon simmering, a process wherein slowly accreting near-$M_{\rm{Ch}}$ WDs develop a large convective core due to energy input from $^{12}$C fusion on timescales of $\sim10^{3}-10^4$ yr before the onset of explosive burning \citep{Wo04, Wu04,PiC08}. Previous studies \citep{Ch08,PiB08} have pointed out that weak nuclear reactions during this phase enhance the level of neutronization in the fuel that will be later consumed in the different regimes of explosive nucleosynthesis. Here, we perform detailed models of slowly accreting WDs with the stellar evolution code \texttt{MESA} \citep{Pa11,Pa13,Pa15}, paying close attention to the impact of carbon simmering on the neutron excess.

This paper is organized as follows. In Section \ref{Neutron}, we provide an overview of the main processes contributing to neutronization in SNe~Ia, and the importance of understanding these processes in the context of observational probes of SN~Ia explosion physics and the pre-SN evolution of their stellar progenitors. In Section \ref{Models}, we outline our simulation scheme and describe our grid of \texttt{MESA} models for accreting WDs. In Section \ref{Results}, we present the main results obtained from our model grid, and in Section \ref{Conclusions}, we summarize our conclusions and suggest directions for future studies.

\section{Neutronization in Type Ia Supernovae} \label{Neutron}

It is commonly accepted that WDs are the end product of most main-sequence stars \citep[see][and references therein]{Al10}. A typical WD is a ${\sim}\, 0.6 \, M_{\odot}$ stellar object made up by a C/O core that encompasses most of its size, surrounded by an thin ${\sim}\, 0.01 \, M_{\odot}\,$He envelope that, in turn, has a shallower ${\sim}\, 10^{-4} \, M_{\odot}\,$H layer on top \citep{Al10}. On the other hand, massive WDs ($M\gtrsim 1.1  M_{\odot}$) are believed to have O/Ne cores. Therefore, the composition of the core and the outer layers strongly depends on the characteristics of the initial star \citep{Rit96}. The specific chemical composition of a WD determines its properties, which can vary after the AGB phase, along the cooling track, via important processes such as convection, phase transitions of the core and gravitational settling of the chemical elements \citep{Al10}. For this abundance differentiation the main role is played by $^{22}$Ne \citep{GB08,Al10} because its neutron excess makes it sink towards the interior as the WD cools. The released gravitational energy by this process influences both the cooling times of WDs \citep{De02} and the properties of SNe Ia \citep{Br11}. 

A critical parameter that controls the synthesis of neutron-rich isotopes in SN Ia explosions is the neutron excess
\begin{equation}
	\eta \, = \, 1 - 2 Y_{e} \, = \, \sum_{i} \dfrac{N_{i} - Z_{i}}{A_{i}} \, X_{i} \, ,
\end{equation}
where $N_{i}$, $A_{i}$ and $Z_{i}$ are the neutron number, the nucleon number and charge of species $i$ with mass fraction $X_{i}$, respectively. The starting value of $\eta$ in the SN Ia progenitor is set by its metallicity. This works as follows. Stars with zero-age main-sequence masses $> 1.3 \, M_{\odot}$ burn hydrogen through the CNO cycle \citep{Th86}. The slowest step is $\rm{^{14}N(p,\gamma)^{15}O}$, which causes all the C, N and O present in the plasma to pile up at $\rm{^{14}N}$. Subsequently, during the hydrostatic He burning, $\rm{^{14}N}$ converts to the neutron-enriched isotope $\rm{^{22}}$Ne through the reaction chain $^{14}\rm{N}(\alpha,\gamma)^{18}\rm{F}(\beta^{+},\nu_e)^{18}\rm{O}(\alpha,\gamma)^{22}\rm{Ne}$.

Because all CNO elements are converted to $^{22}$Ne during He burning, there is a linear relationship between the metallicity of a main sequence star and the neutron excess in the WD it eventually produces. Indeed, \citet{Ti03} found that this process relates the neutron excess of the WD and its progenitor metallicity via $\eta = 0.101 Z$, where $Z$ refers to the mass fraction of CNO elements, resulting in a value for solar metallicity material of $\eta_{\odot} = 1.4 \times 10^{-3}$. Gravitational settling of $^{22}\rm{Ne}$ might enhance the relative neutronization in the core, but only at the expense of shallower material from the outer layers \citep{PiC08}.

\subsection{Neutron production during carbon simmering} \label{Simmering}

\placefigure{f1}
\begin{figure*}
\centering
\includegraphics[scale=0.75]{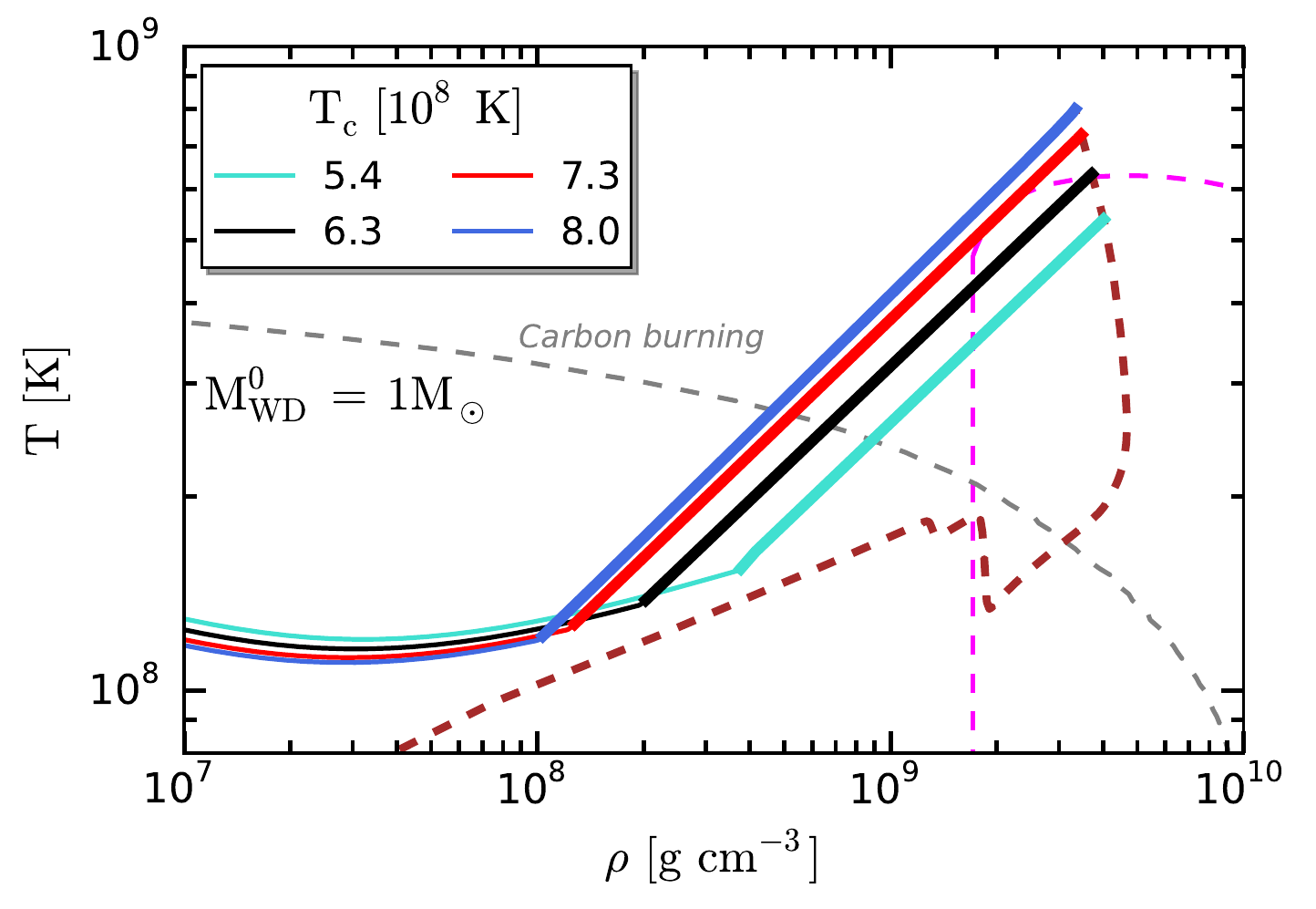}
\caption{Temperature versus density profiles taken from our fiducial model (Section \ref{Fiducial}), presented analogously to Figure 1 from \citet{PiB08}. Each profile represents a snapshot in time as the central temperature increases and the convective region grows. The convective region of each profile is represented with thick lines. The dashed, brown line tracks the central density and temperature over time, showing how the central density decreases as the central temperature increases during simmering. The two sharp drops at $\log(\rho_{\mathrm{c}}/\g\cm^{-3}) \approx 9.1-9.2$ correspond to neutrino losses in the $^{23}$Na--$^{23}$Ne and $^{25}$Mg--$^{25}$Na Urca shells, as explained in Section \ref{Urca} and shown in Figure \ref{TcvsRhoc_Urca}. 
The dashed, magenta line shows where the heating timescale and $^{23}$Na electron-capture timescale are equal; at lower densities/higher temperatures, electron captures on $^{23}$Na are frozen out. The dashed, gray line is an approximate C-ignition curve from \texttt{MESA} that considers a 100\% carbon composition in the core.}
\label{RhoT}
\vspace{0.25 cm}
\end{figure*}

This relation between $\eta$ and $Z$ can subsequently be modified by carbon simmering \citep{PiB08,Ch08}, and we summarize the main features of this process in Figure \ref{RhoT}. Carbon ignition in the core of a WD takes places through the channels $\rm{^{12}C(^{12}C,\alpha)^{20}Ne}$ and $\rm{^{12}C(^{12}C,p)^{23}Na}\,$ with a branching ratio 0.56/0.44 for $T < 10^{9} \, \rm{K}$ \citep{Ca88} when the heat from these nuclear reactions  surpasses the neutrino cooling. This burning regime \citep{No84}, which starts at the gray, dashed line in Figure~\ref{RhoT}, marks the onset of simmering. The central conditions then trace out the rising dashed, brown line as the star heats and decreases slightly in density. At the same time, a convective region grows outward \citep{Wo04,Wu04}, shown at four different epochs with thick, solid lines. This convection encompasses $\, \sim 1 \, M_{\odot}$ during a period of $\, \sim10^{3}-10^4 \,$yr before the final thermonuclear runaway at a central temperature of $T_c\approx8\times10^8\,{\rm K}$ and the explosion as a Type Ia~SN \citep{PiC08}.

During carbon simmering, the protons produced by the $\,\rm{^{12}C(^{12}C,p)^{23}Na}\,$ reaction capture onto $^{12}$C, producing $^{13}$N. Subsequently, the electron-capture reactions $\,^{13}$N($\rm{e^{-}}$, $\rm{\nu_{e}}$)$^{13}$C$\,$ \citep[discussed in detail in Section 2.2 of][]{Ch08} and $\,^{23}$Na($\rm{e^{-}}$, $\rm{\nu_{e}}$)$^{23}$Ne$\,$ \citep{Ch08,PiB08}  produce an enhancement in the neutronization of the core. These reactions consume the products of carbon fusion, so this increase in $\eta$ is directly related to the amount of carbon consumed prior to the explosion. This proceeds until sufficiently high temperatures or low densities are reached such that timescale for the $^{23}$Na electron captures becomes longer than the heating timescale. (The location where these timescales are equal is shown as a magenta, dashed line in Figure \ref{RhoT}.)  Additionally, as we find here, $\,^{23}$Ne carried into lower density regions of the convection zone can be converted back to $^{23}$Na by beta decay.  These nuclear processes determine the final composition and properties of the ejected material \citep{Iw99} and are crucial to obtain synthetic spectra \citep{Br00}.

Using these basic arguments, \citet{PiB08} semi-analytically calculated the amount of carbon consumed during simmering to estimate that the increase in the neutron excess should be $\Delta \eta \, {\sim} \, 10^{-3}\,$ with an upper bound of 0.93$\, \eta_{\odot}$ known as the ``simmering limit''. Such a floor to the neutron excess is important to identify because it should be present in any SN Ia progenitor that went through a simmering phase, regardless of how low the progenitor's metallicity is. Using a more detailed nuclear network, but only focusing on the central conditions of the convective zone, \citet{Ch08} predicted a decrement in the mean number of electrons per baryon of $|\Delta Y_{e}| = 2.7-6.3 \times 10^{-4}$, which corresponds to $\Delta \eta \,\, {\approx} \,\, 5.4-13\times10^{-4}\,$. Although both these works found similar levels of neutronization, they also made strong simplifications, and this is an important motivation for revisiting these results here.

\subsection{Urca-process cooling} \label{Urca}

Weak reactions can also affect the thermal state of the WD.  An Urca pair consists of two nuclei $(Z,A)$ and $(Z-1, A)$ that are connected by electron-capture $(Z,A) + e^- \to (Z-1,A) + \nu_e$ and beta-decay $(Z-1,A) \to (Z,A) + e^- + \bar{\nu}_e$.  Below a threshold density $\rho_{\mathrm{th}}$ the beta-decay reaction is favored and above it the electron-capture reaction is favored.  Near this threshold density, both reactions occur at a significant rate, and since each produces a neutrino that then free-streams from the star, this has the net effect of cooling the plasma \citep{Gamow41}.

\begin{figure}
\includegraphics[scale=0.61]{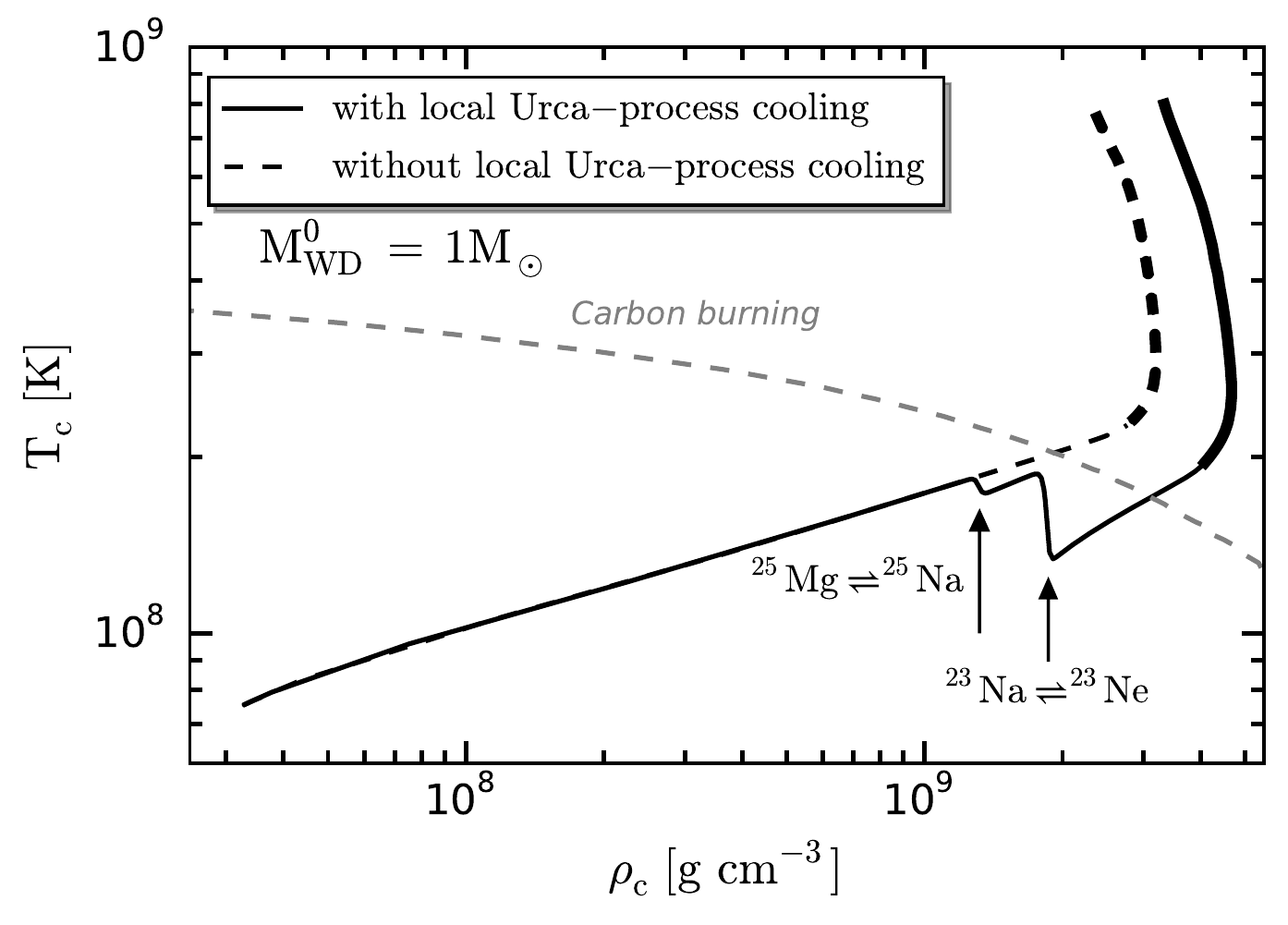}
\caption{A comparison of the evolutionary tracks for the central density and temperature in our fiducial model (Section \ref{Fiducial}) with (black line) and without (dashed line) the effects of the $^{23}$Na--$^{23}$Ne and $^{25}$Mg--$^{25}$Na Urca pairs (see Section \ref{Urca}). The evolution during the simmering phase is denoted with thick lines. The gray, dashed line is an approximate C-ignition curve from \texttt{MESA} that considers a 100\% carbon composition in the core. }
\label{TcvsRhoc_Urca}
\end{figure}

As the WD is compressed, its density increases above the threshold density of numerous Urca pairs. For the compositions and densities of our WD models, the two most important Urca pairs are $^{25}$Mg--$^{25}$Na (with $\log(\rho_{\rm{th}}/\g\cm^{-3}) \approx 9.1$) and $^{23}$Na--$^{23}$Ne (with $\log(\rho_{\rm{th}}/\g\cm^{-3}) \approx 9.2$) \citep{Iben78}. These threshold densities are below the density at which carbon ignition occurs, and hence in the central parts of the WD \textit{this local Urca-process cooling occurs before the simmering phase begins}.  This effect was discussed in the context of accreting C/O cores by \citet{Paczy73}, but is often not included in progenitor models.\footnote{This effect was included in a recent study by \citet{Den15}, who used a large nuclear network that incorporated new weak rate tabulations from \citet{Toki13}.} We make use of new capabilities of \texttt{MESA} that allow these processes to be easily included \citep{Pa15}. As shown in Figure \ref{TcvsRhoc_Urca}, additional cooling shifts the point at which carbon ignition occurs to higher densities.  The specific energy loss rate due to Urca-process neutrinos scales $\propto T^4$ \citep{Tsuruta70}, so this effect is most pronounced in hotter WDs (those with short cooling ages).

\placefigure{f2}

After carbon ignition occurs and the simmering phase begins, the Urca process continues to operate as convection mixes material from regions where it has electron-captured into regions where it will beta-decay and vice-versa.  This \textit{convective} Urca process and its effects have been an object of considerable study \citep[e.g.,][]{Paczy72,Bruenn73,Shaviv77,Barkat90,Le05}.  We allow for the operation of the convective Urca process in our \texttt{MESA} models, inasmuch as we incorporate appropriate weak rates and allow composition to mix throughout the convective zone.  However, limitations imposed by the temporal and spatial averaging that enter into a formulation of 1D mixing-length theory do not allow us to self-consistently treat the effects of the Urca process on the convection itself. 
In some of our models, in particular those with the solar or super-solar metallicities and hence the highest abundances of $^{25}$Mg and $^{23}$Na, we observe that, when the convective zone first reaches the Urca shell, the former splits in two and remains split for the remainder of the calculation.  It seems likely this behavior is a manifestation of these limitations, so when we report our results in Tables~\ref{Table_0Gyr}-\ref{Table_10Gyr}, we mark these models with the note ``Convection zone splits during simmering''.  The development of a model able to fully incorporate the interaction of convection and the Urca process is beyond the scope of this work and will likely require multi-dimensional hydrodynamics simulations \citep[e.g.,][]{St06}.  Given the existing uncertainties, \citet{Den15} explored the possible effects of the convective Urca process in \texttt{MESA} models by employing a series of mixing assumptions, such as limiting the mass of the convective core to the mass coordinate of the  $^{23}$Na--$^{23}$Ne Urca shell.  Future work could employ a similar approach to explore the potential effects of the convective Urca process on neutronization.

\section{White Dwarf Models} \label{Models}

Motivated by the discussion above, we next explore the impact of the neutron-rich isotopes at WD formation and during carbon simmering using \texttt{MESA}\footnote{\underline{\texttt{\url{http://mesa.sourceforge.net/index.html}}}\color{black}} \citep{Pa11,Pa13,Pa15}. We create WDs with five different metallicities: $Z/Z_{\odot} = 0.01, 0.10, 0.33, 1.00, 2.79$\footnote{Our intention was to create a 3$Z_{\odot}$ star. However, \texttt{MESA} presents several convergence problems for this high metallicity during the AGB phase and a 2.79$Z_{\odot}$ WD was created instead.}, where $Z_{\odot} = 0.014$ \citep{As09}. In each case, we start from 4.5$M_{\odot}$ ZAMS-models by using the inlists from the suite case \texttt{make\_co\_wd}, which makes a protostar go through the MS until the AGB thermal pulses and then reveals its C/O core. These models are stopped when the total luminosity reaches $\log{L/L_{\odot} = -0.5}$. \texttt{MESA} presents convergence problems due to the unstable He shell burning on accreting WDs \citep{Sh09}, so we artificially remove the H/He shallower envelope via a negative accretion rate. The resulting WDs have $\log{L/L_{\odot} \approx -1.4}$. Then, we rescale the initial masses of our WDs to 0.70, 0.85 and 1.00$M_{\odot}$ without changing the chemical composition as a function of the Lagrangian mass coordinate.

To evaluate the effect of cooling times in the properties of WDs \citep{Le06,Al10}, we let our stars cool for 1 and 10 Gyr, ages that are consistent with the spread for the delay-time distribution (DTD) of SNe Ia \citep[${\sim} \,$40 Myr--10 Gyr,][]{Ma12,Ma14}.  We do not account for residual heating by the external H/He envelope \citep{Al10}, as this material has already been removed in our models.  We also do not include the effects of diffusion, sedimentation, or crystallization, as the development of \texttt{MESA}'s treatment of these processes is ongoing.  With these caveats in mind, we classify our WDs as ``hot'' (no cooling applied), ``warm'' (1 Gyr) and ``cold'' (10 Gyr).

We use these 45 WDs (five metallicities, three masses, and three cooling ages) as an input for our simmering  \texttt{MESA} inlists, based on the suite case \texttt{wd\_ignite}, which models the accretion in the Type Ia SNe SD channel by considering a C/O WD, a uniform, pure C/O accretion and a stopping condition such that the total luminosity from the nuclear reactions reaches $10^{8} \, L_{\odot}$.  We use a nuclear network consisting of 48 isotopes, shown in Table \ref{Table_isotopes}, and the reactions linking them.  This is the main difference between the present study and the one performed by \citet{Ch14}, who also examined the properties of accreting C/O WDs, but used a more limited network. We use a version of \texttt{MESA} based on release 7624, but modified to incorporate a rate for the $\,^{13}$N($\rm{e^{-}}$, $\rm{\nu_{e}}$)$^{13}$C$\,$ reaction that is appropriate for the high density conditions of a WD interior.  We motivate and describe our modifications in Appendix \ref{AppA}.

\begin{table}
\begin{center}
\caption{Nuclear network used in our calculations.\label{Table_isotopes}}	
\begin{tabular}{crcr}
\tableline\tableline
\noalign{\smallskip}
Isotope & $A$ & Isotope & $A$ \\
\noalign{\smallskip}
\tableline
\noalign{\smallskip}
n & 1 & O & 14--18\\
H & 1--2 & F & 17--19\\
He & 3--4 & Ne & 18--24\\
Li & 7 & Na & 21--25\\
Be & 7 & Mg & 23--26\\
Be & 9--10 & Al & 25--27\\
B & 8 & Si & 27--28\\
C & 12--13 & P & 30--31\\
N & 13--15 & S & 31--32\\
\noalign{\smallskip}
\tableline
\end{tabular}
\end{center}
\end{table}

For the accreted material, we consider uniform accretion with three different rates, $10^{-6}$, $10^{-7}$ and $5 \times 10^{-8} \, M_{\odot}\,{\rm yr^{-1}}$, which yield accretion ages $\sim 10^{6} \, $yr that agree with the literature \citep[]{Ha96, Ha04}. The chemical abundances of the accretion are set equal to the initial surface composition of each WD. This makes a total of 135 different models whose results are presented in Section \ref{Global_results}. In order to achieve a higher spatial and temporal resolution during the Urca-process cooling and carbon simmering phases, we stop our accreting models when the WD mass reaches 1.3 $M_{\odot}$, which corresponds to central densities below the Urca cooling densities discussed in Section \ref{Urca}.  We then continue the models with controls that incorporate timestep limits based on the the variation of the central density and temperature. To confirm that our models are converged, we perform runs with increased temporal and spatial resolution. To corroborate that our results are insensitive  to the treatment of the outer boundary of the central convection zone, we execute a run with overshooting. We verify that the quantities of interest are unchanged and refer the reader to Appendix \ref{AppB} for more detailed information.

\begin{figure}
\includegraphics[scale=0.63]{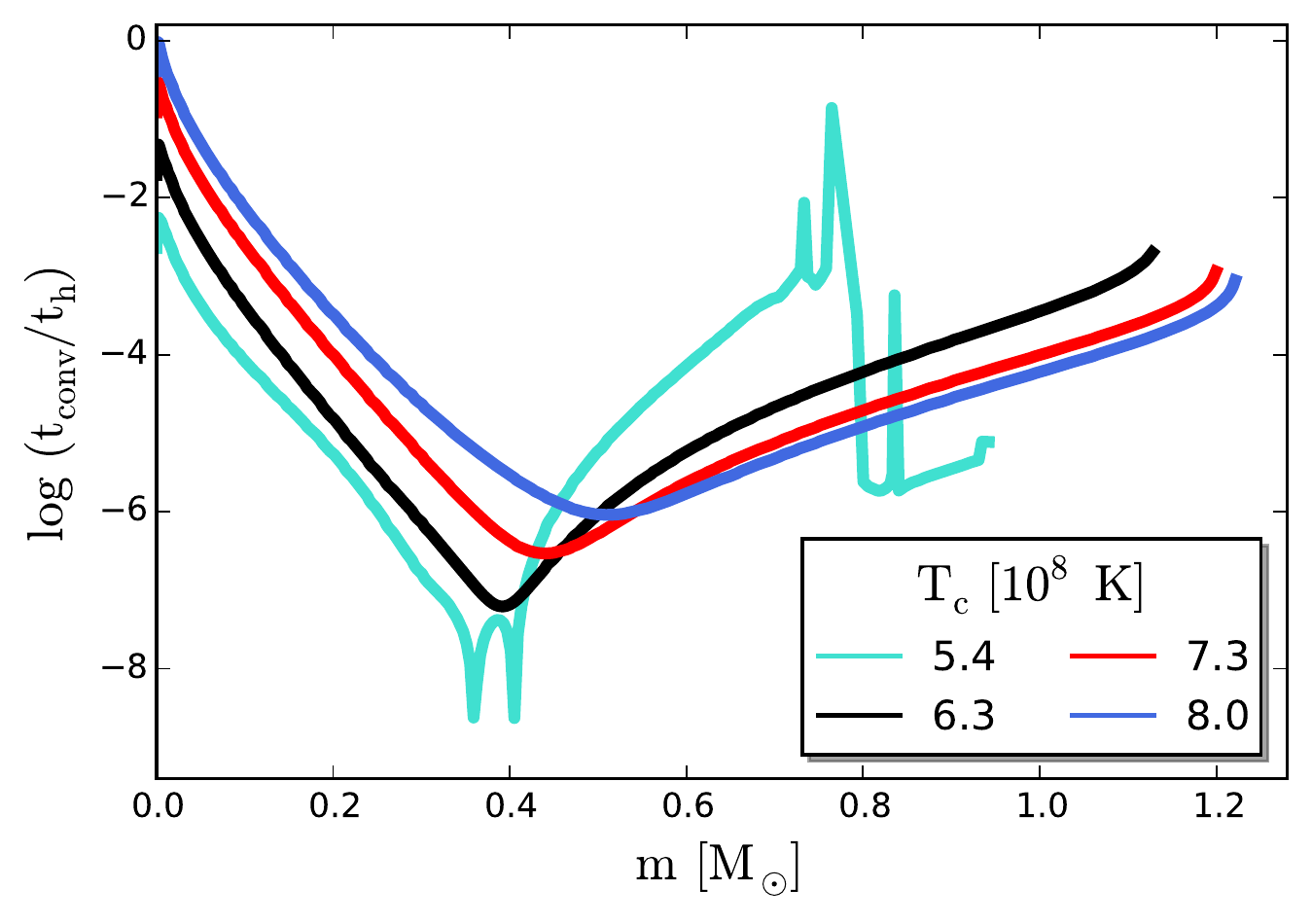}
\caption{Profiles of the ratio between the convective and the heating time scales versus the Lagrangian mass in the growing convective region for our fiducial model (Section \ref{Fiducial}). The convective overturn timescale $t_{\rm{conv}}$ gets comparable to $t_{h}$ at the center of the WD right before the final thermonuclear runaway as shown by the blue curve. Various nuclear reactions with rates $\lambda$ should freeze out when $t_{h} < \lambda^{-1}$.}
\label{tcth}
\end{figure}

For the stopping condition of our inlists, we choose the one derived by \citet{Wo04}, and broadly discussed in \citet{Ch08}, \citet{PiB08} and \citet{PiC08}, which estimates that simmering should end when dynamical burning is triggered. This requires $T_{\rm{c}} \approx 8 \times 10^{8} \, \rm{K}$, i.e., $\log(T_{\rm{c}}/\mathrm{K}) \approx 8.9$. In turn, \citet{PiB08} argued that the final thermonuclear runaway should ensue when the heating time scale $t_{h} = c_{\rm{p}} T_{\rm{c}}/\epsilon$ (where $c_{\rm{p}}$ is the specific heat of the liquid ions) gets comparable to the dynamical time scale $t_{\rm{dyn}} \equiv (G \rho_{\rm{c}})^{-1/2} \, {\sim} \, 1 \, \rm{s}$. Our conclusion is that, in general, $t_{h} \gtrsim 10 t_{\rm{dyn}}$ when $T_{\rm{c}} \approx 8 \times 10^{8} \, \rm{K}$, so that $t_{h} \, {\sim} \, t_{\rm{dyn}}$ does not hold. Figure \ref{tcth} shows that simmering ends when the heating time scale approaches the convective time scale $t_{h} \leq t_{\rm{conv}}$ in the core of the WD \citep{PiC08}. Here, $t_{\rm{conv}} = \rm{min}$$\{H_p,R_{\rm{conv}}\}/v_{\rm{conv}}$, where $v_{\rm{conv}}$ is the convective velocity, $R_{\rm{conv}}$ the extent of the convective zone and $H_p$ the pressure scale height, which \texttt{MESA} calculates as $H_{p} = \rm{min}$$\{ P/(g \rho) , \sqrt{P/G}/\rho \}$.

\placefigure{f3}
\vspace{0.1 cm}

\section{Results} \label{Results}

We next summarize the main results of our survey of simmering WD models. We begin in Section \ref{Fiducial} by focusing on a fiducial model, a 1 $M_{\odot}$, solar-metallicity WD with an accretion rate of $10^{-7} \, M_{\odot}\,{\rm yr^{-1}}$. This is used to compare and contrast with our large grid of models in Section \ref{Global_results}.

\subsection{Fiducial model} \label{Fiducial}

\begin{figure}
\includegraphics[scale=0.61]{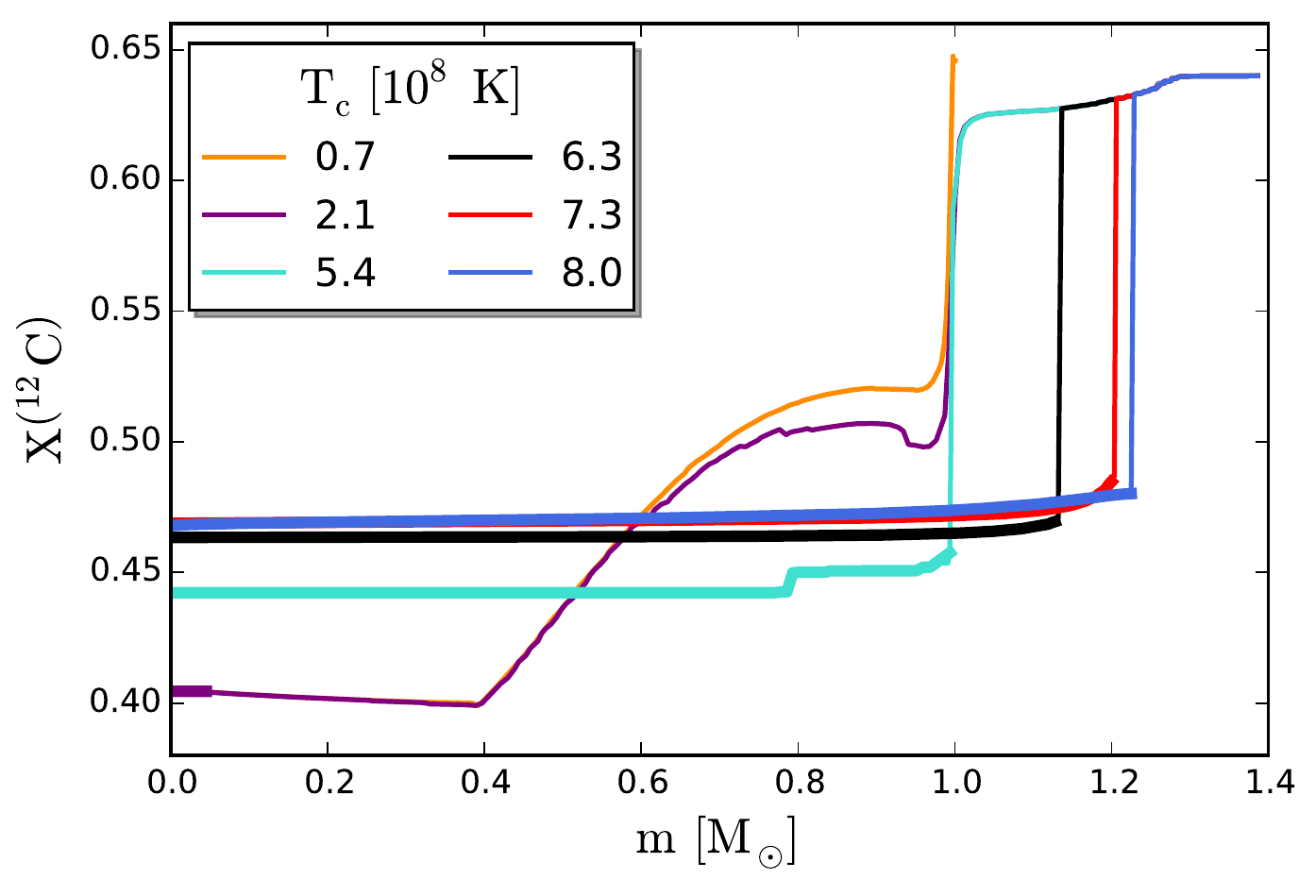}
\includegraphics[scale=0.61]{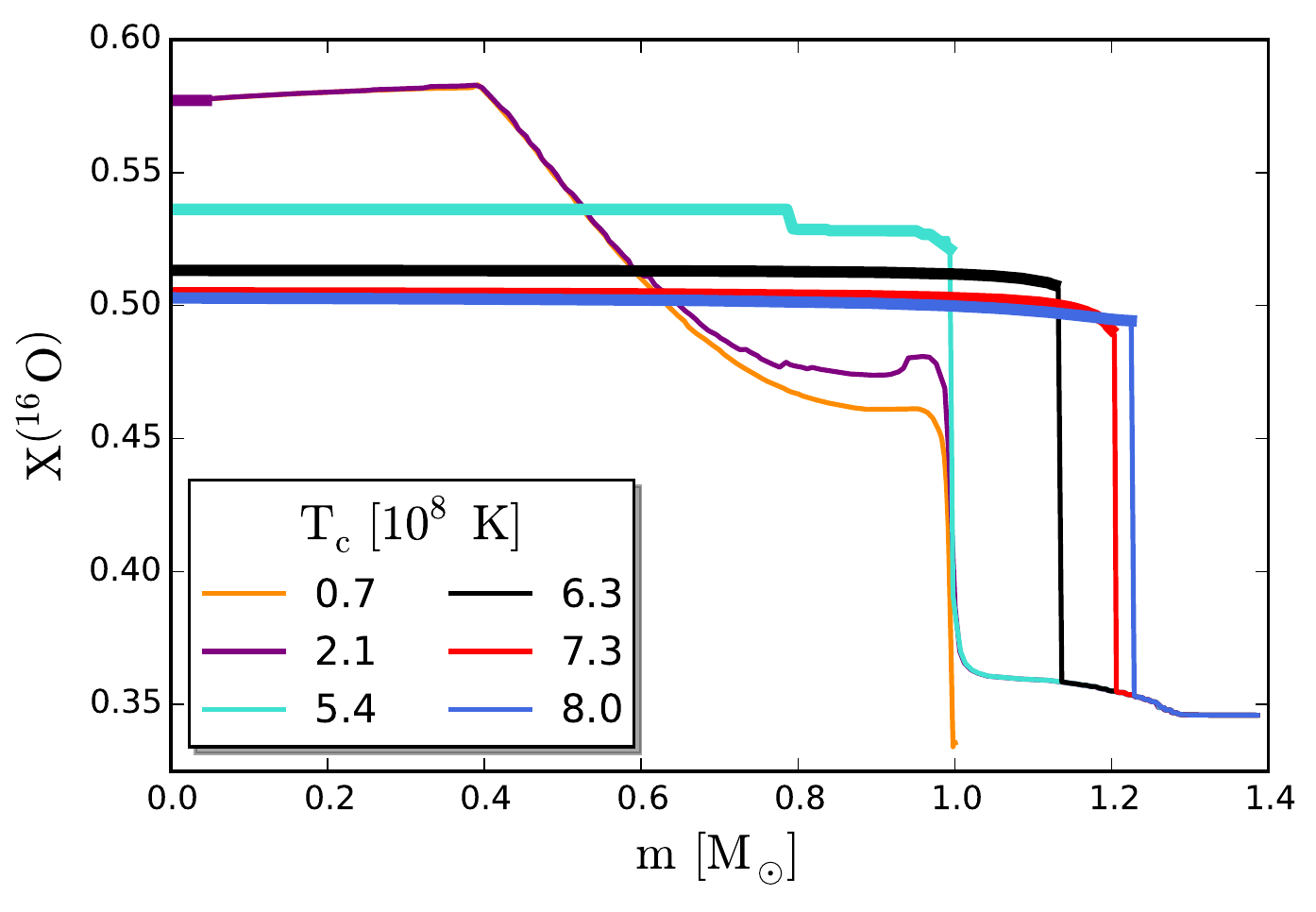}
\includegraphics[scale=0.59]{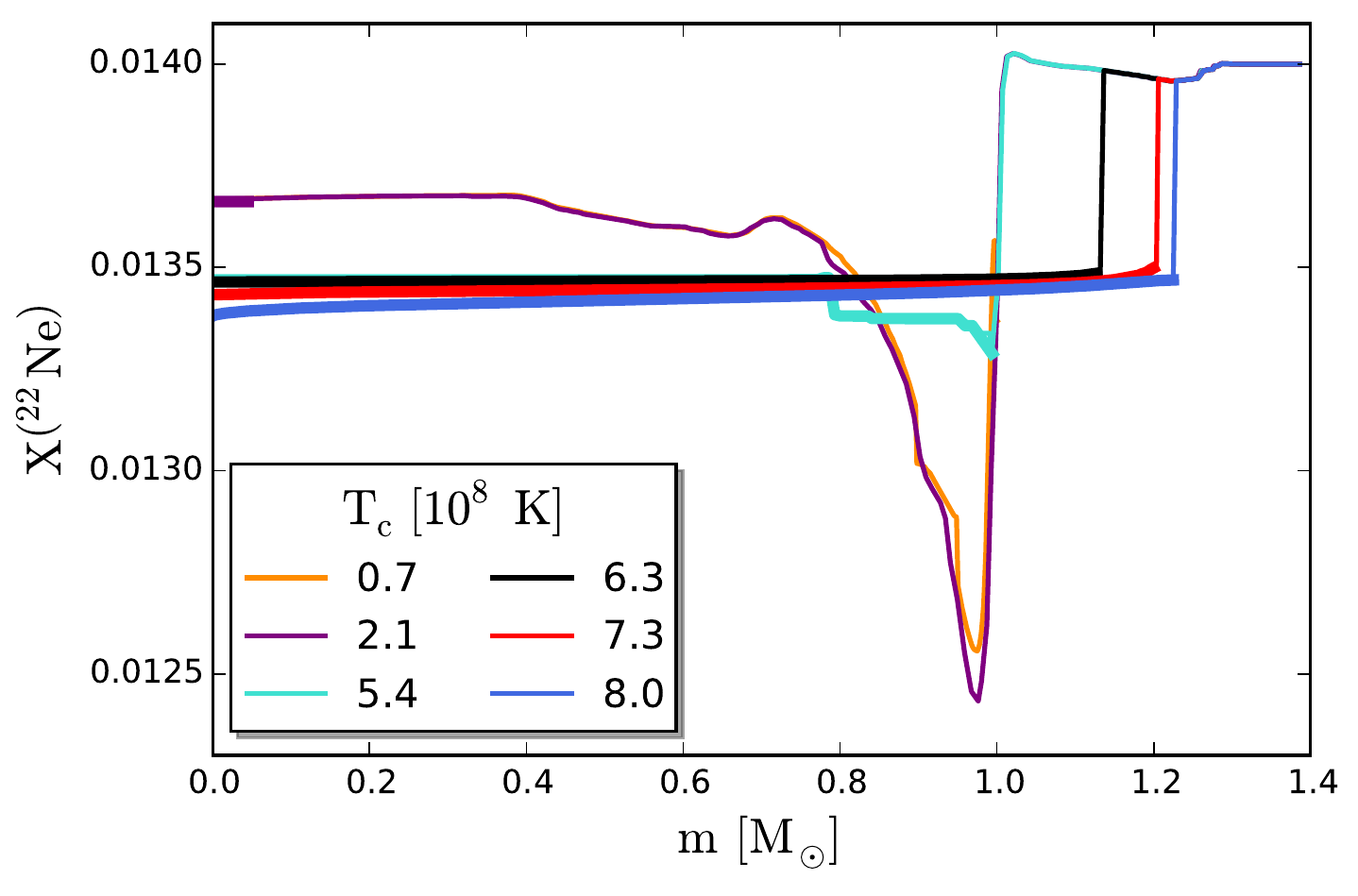}
\caption{Abundance profiles of $^{12}$C (top), $^{16}$O (middle) and $^{22}$Ne (bottom) in our fiducial model. The orange curve represents the initial model, while the purple one corresponds to the onset of carbon simmering. The convective region of each profile is depicted with thick lines.}
\label{c12o16ne22}
\end{figure}

In Figure \ref{c12o16ne22}, we show the evolution of the chemical profiles of $^{12}$C, $^{16}$O and $^{22}$Ne for our fiducial model during the different stages of the accretion process and through the simmering phase. We have labeled our curves at different time steps with the corresponding central temperature because of our stopping condition \citep{Wo04}. After carbon simmering, all the chemical profiles become homogeneous within the convective core (shown by the thick, fairly flat regions of the profiles). In turn, the accreted material eventually gets mixed into the core when the edge of this convective region reaches the initial mass of the star, which is why the carbon fraction increases at the center (carbon-rich material is mixed in rate higher than the consumption by carbon burning). The last profiles show a clear distinction between convection, which encompasses $\approx 90 \%$ of the star by mass, and the outer, non-convective regions of the WD.

\placefigure{f4a}
\placefigure{f4b}
\placefigure{f4c}

\placefigure{f5}
\begin{figure}
\includegraphics[scale=0.63]{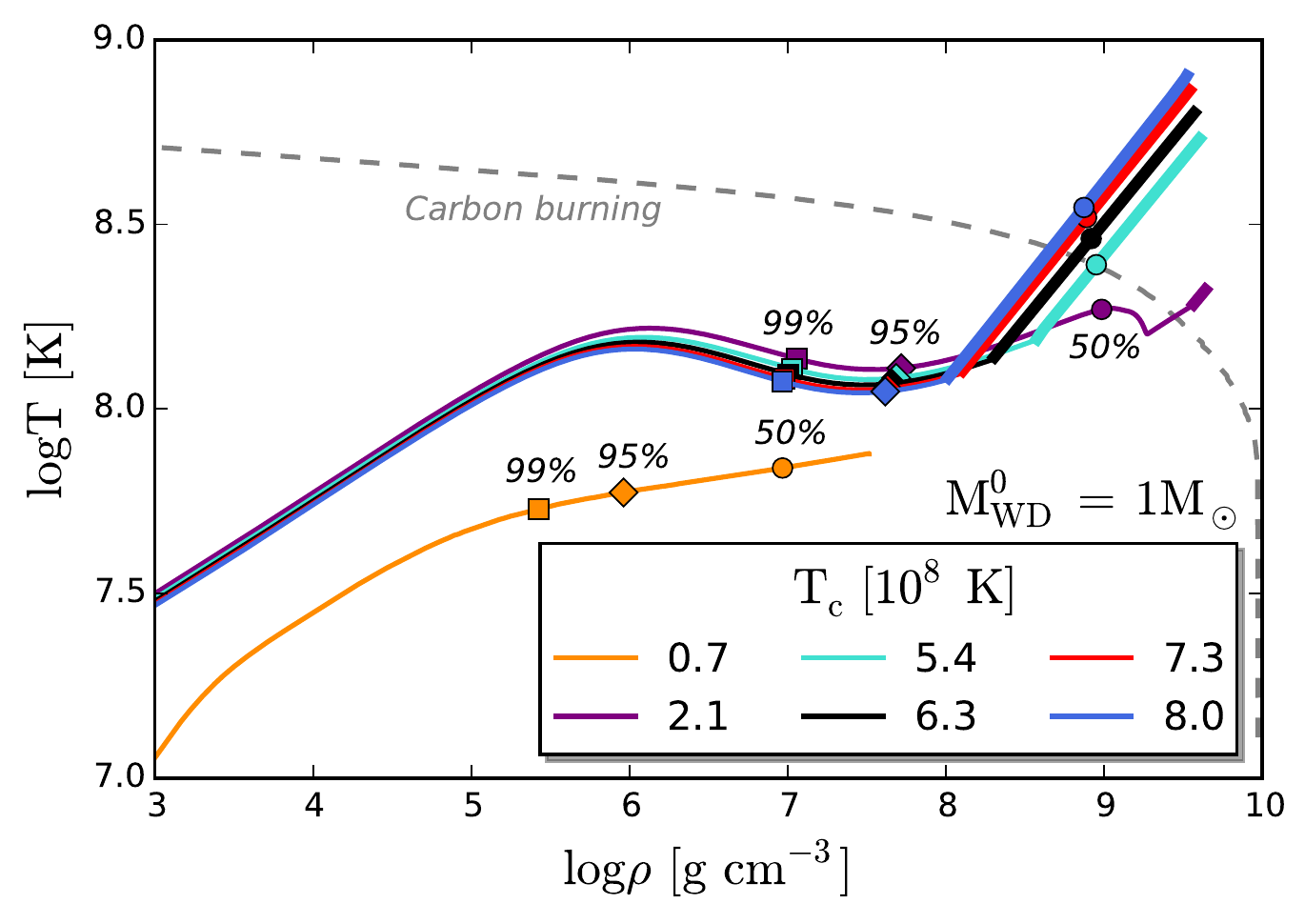}
\caption{Temperature versus density profiles during several stages of the stellar evolution in our fiducial model. The color legend is the same as the one of Figure \ref{c12o16ne22}, whereas the gray, dashed line is an approximate C-ignition curve from \texttt{MESA} that considers a 100\% carbon composition in the core, which is why the purple profile does not exactly match it. Finally, some points encompassing fractions of the stellar mass are depicted along each of the curves.}
\label{EOS}
\end{figure}

\placefigure{f6}
\begin{figure}
\includegraphics[scale=0.61]{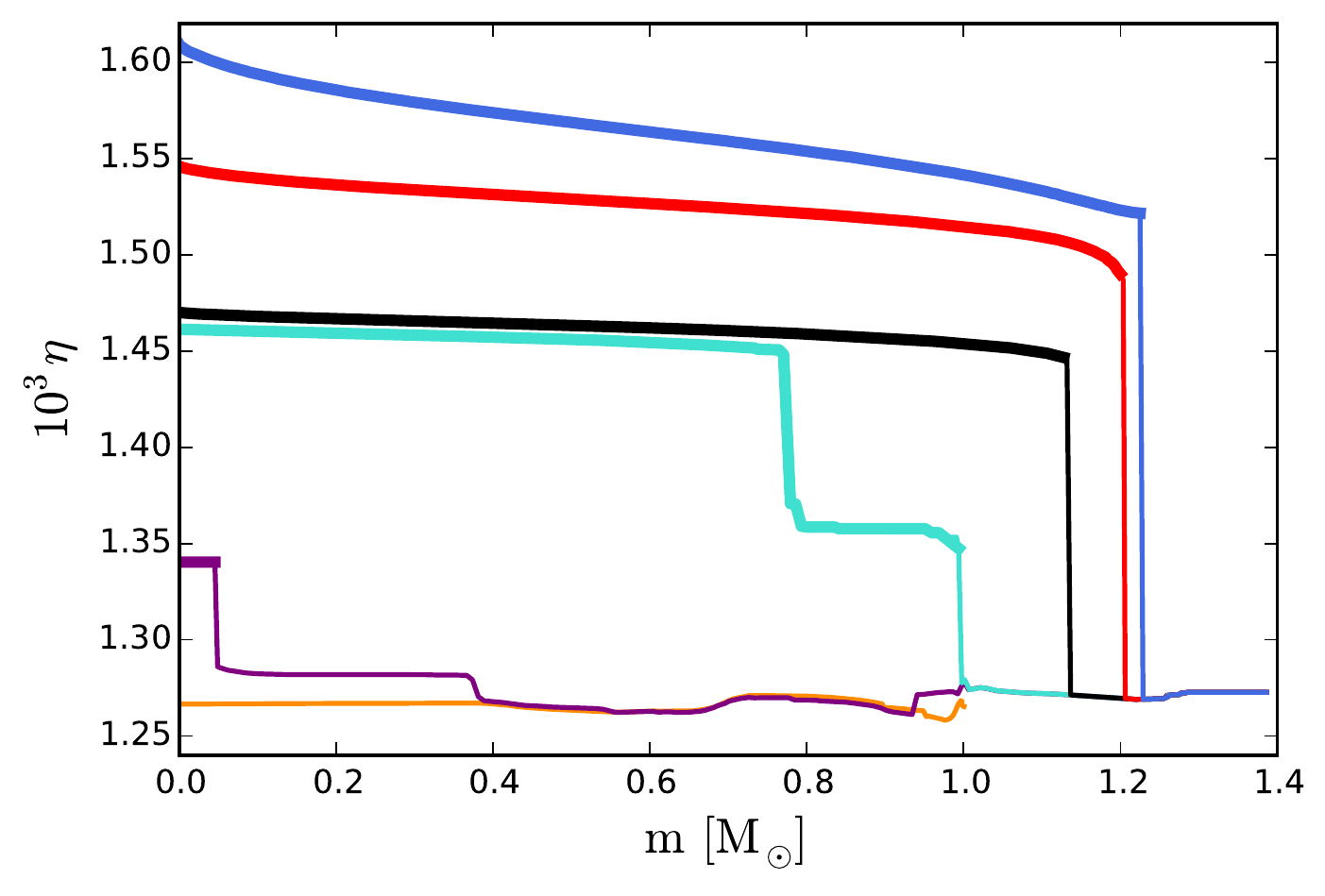}
\caption{Neutron excess profiles as a function of the Lagrangian mass for the same series of snapshots as shown in Figure \ref{c12o16ne22}.}
\label{eta}
\end{figure}

\placefigure{f7}
\begin{figure}
\includegraphics[scale=0.61]{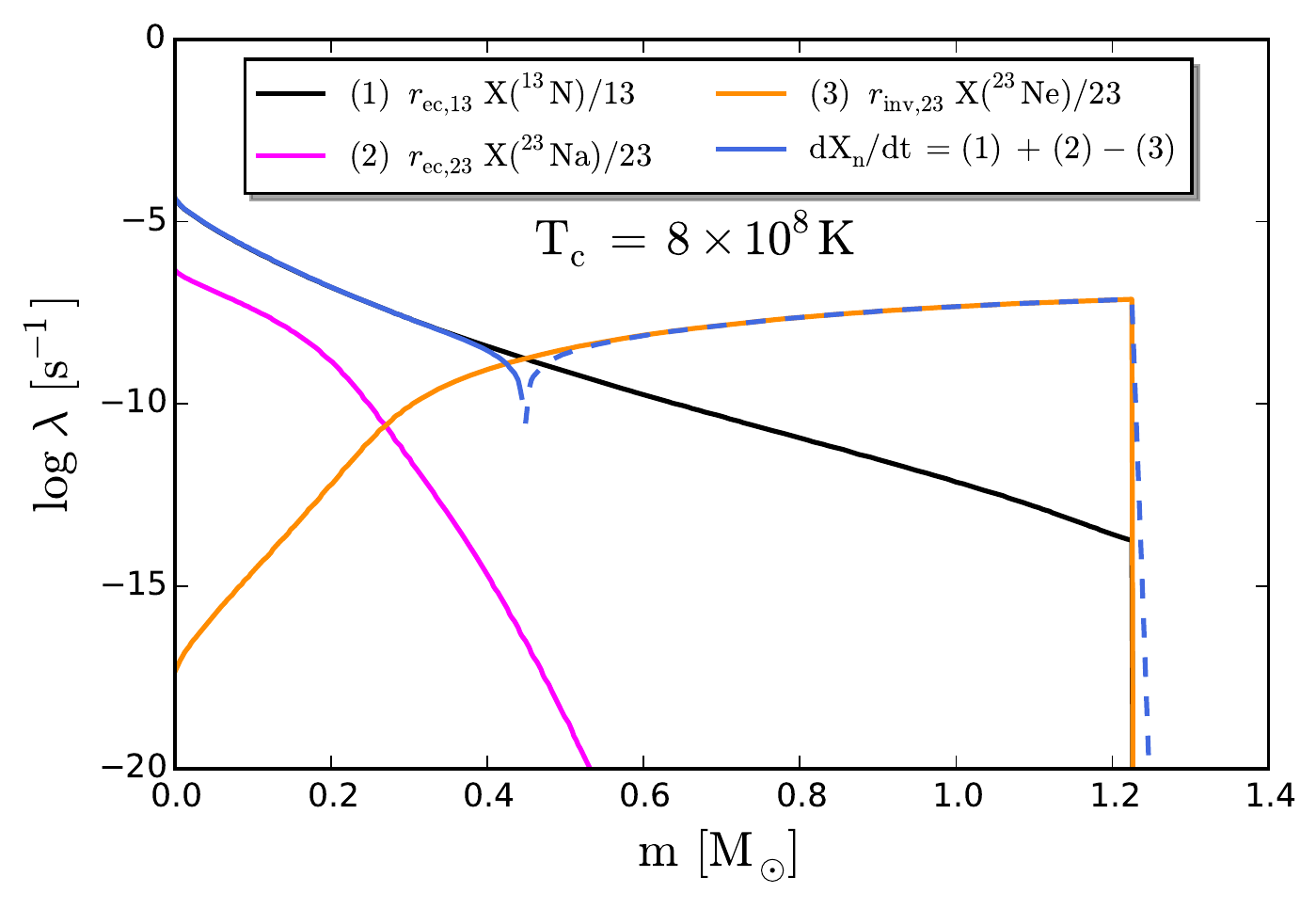}
\caption{Profile of the variation of the neutron fraction $dX_n/dt$ for $T_{\rm{c}} = 8 \times 10^{8} \, \rm{K}$ (blue) and the rates $\lambda$ of the three weak reactions involved. The dashed line indicates the region where it is negative. The black and the magenta lines refer to, respectively, the electron capture reactions $\,^{13}$N($\rm{e^{-}}$,$\rm{\nu_{e}}$)$^{13}$C$\,$ and $\,^{23}$Na($\rm{e^{-}}$,$\rm{\nu_{e}}$)$^{23}$Ne$\,$. Finally, the orange line is the beta decay $\,^{23}$Ne($\rm{\nu_{e}}$,$\rm{e^{-}}$)$^{23}$Na$\,$ whose dominance in the outer, lower-density regions explains why the increase in the neutron excess is smaller than the one predicted by \citet{PiB08} and \citet{Ch08}.}
\label{weak_rates}
\end{figure}

\placefigure{f8}
\begin{figure}
\includegraphics[scale=0.61]{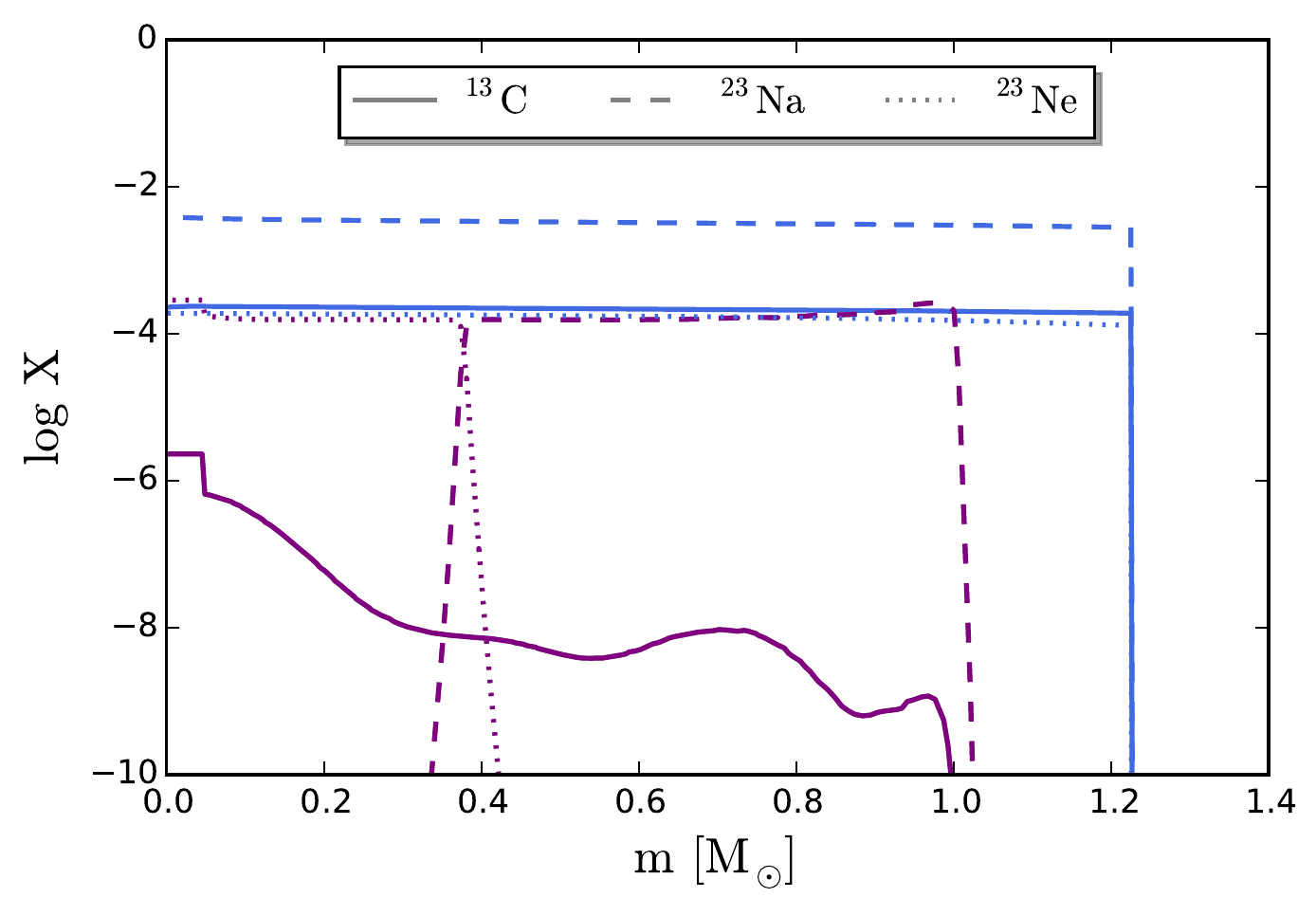}
\caption{Abundance profiles of $^{13}$C, $^{23}$Na and $^{23}$Ne in our fiducial model at the onset of simmering ($T_{\rm{c}} = 2.1 \times 10^{8} \, \rm{K}$; purple lines) and the end of our calculation ($T_{\rm{c}} = 8 \times 10^{8} \, \rm{K}$; blue lines).  During simmering, the convection zone is fully mixed, allowing $^{23}$Ne to be converted back to $^{23}$Na when it is transported below the threshold density.}
\label{c13na23ne23}
\end{figure}

Figure \ref{EOS} shows the $\log T - \log \rho$ profiles for the fiducial model. Initially, the hot, accreted material increases the effective temperature of the WD, while the interior of the star remains unchanged. After ${\sim}\, 10^{3-4}$ years, the temperature gradient steepens due to the energy lost via neutrinos \citep[$\propto T^{3}$,][]{Ch14}, so that a temperature inversion arises in the outer regions of the WD. This is critical because, for high accretion rates and cold WDs, the outer layers will be hotter than the core and off-center ignitions might take place \citep{Ch14}. Finally, there is a change in the thermal structure of the star after the onset of simmering. Since convection is very efficient in the core given the high thermal conduction timescale $\sim 10^{6} \,$yr, the convective profile is nearly an adiabat \citep{Pi08}.

Figure \ref{eta} shows the neutron excess as a function of depth. This starts relatively constant with depth at a value of $\eta\approx1.25-1.3\times10^{-3}$ set by the progenitor metallicity. Then, as the simmering proceeds, a region with an increased neutron excess is seen to grow out in mass. At the onset of thermonuclear runaway, the central neutron excess is enhanced by an amount $\approx 3 \times 10^{-4}$, so that $Y_{e}$ is reduced by $\approx 1.5 \times 10^{-4}$. This is smaller than the decrement within the convective zone at the center $|\Delta Y_{e}| = 2.7-6.3 \times 10^{-4}$ predicted by \citet{Ch08}, as well as than the maximum neutronization estimate $|\Delta Y_{e}| \approx 6 \times 10^{-4}$ calculated by \citet{PiB08}. The reason for this discrepancy is that we have resolved the entire convective zone at each time and the range of densities encompassed by it. The electron captures are very sensitive to density, and thus the outer, lower-density regions do not experience the same level of electron captures and corresponding neutronization. This can be appreciated in Figures \ref{weak_rates} and \ref{c13na23ne23}. As we find here, $^{23}$Ne can be converted back to $^{23}$Na when it is carried into the portion of the convection zone below the threshold density. In contrast, both \citet{PiB08} and \citet{Ch08} focused on the central, highest-density conditions for deriving rates, and thus overestimated the amount of neutronization.

\subsection{Cooled models and global results} \label{Global_results}

\placefigure{f9a}
\placefigure{f9b}
\placefigure{f9c}
\begin{figure}
\includegraphics[scale=0.6]{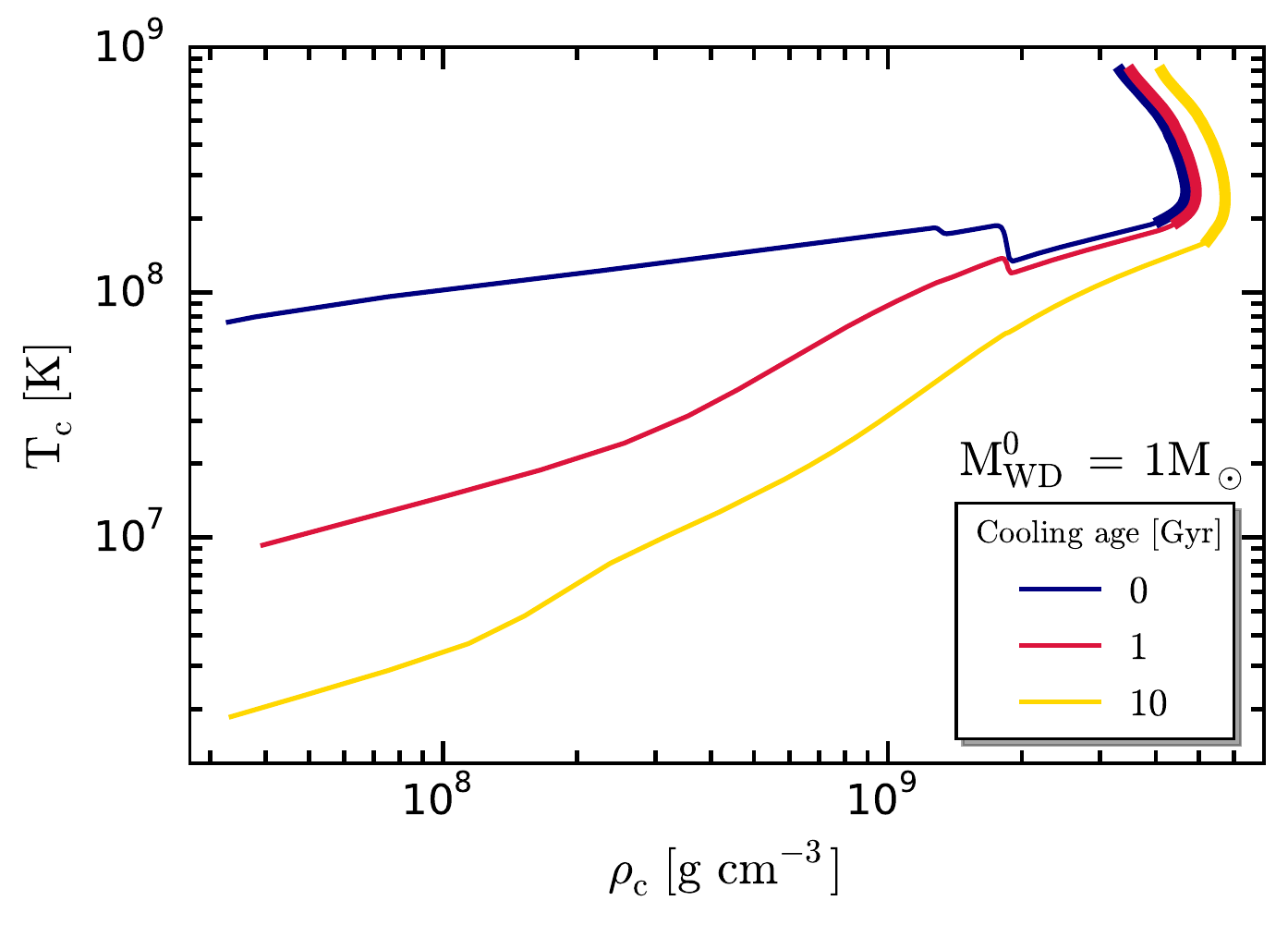}
\includegraphics[scale=0.61]{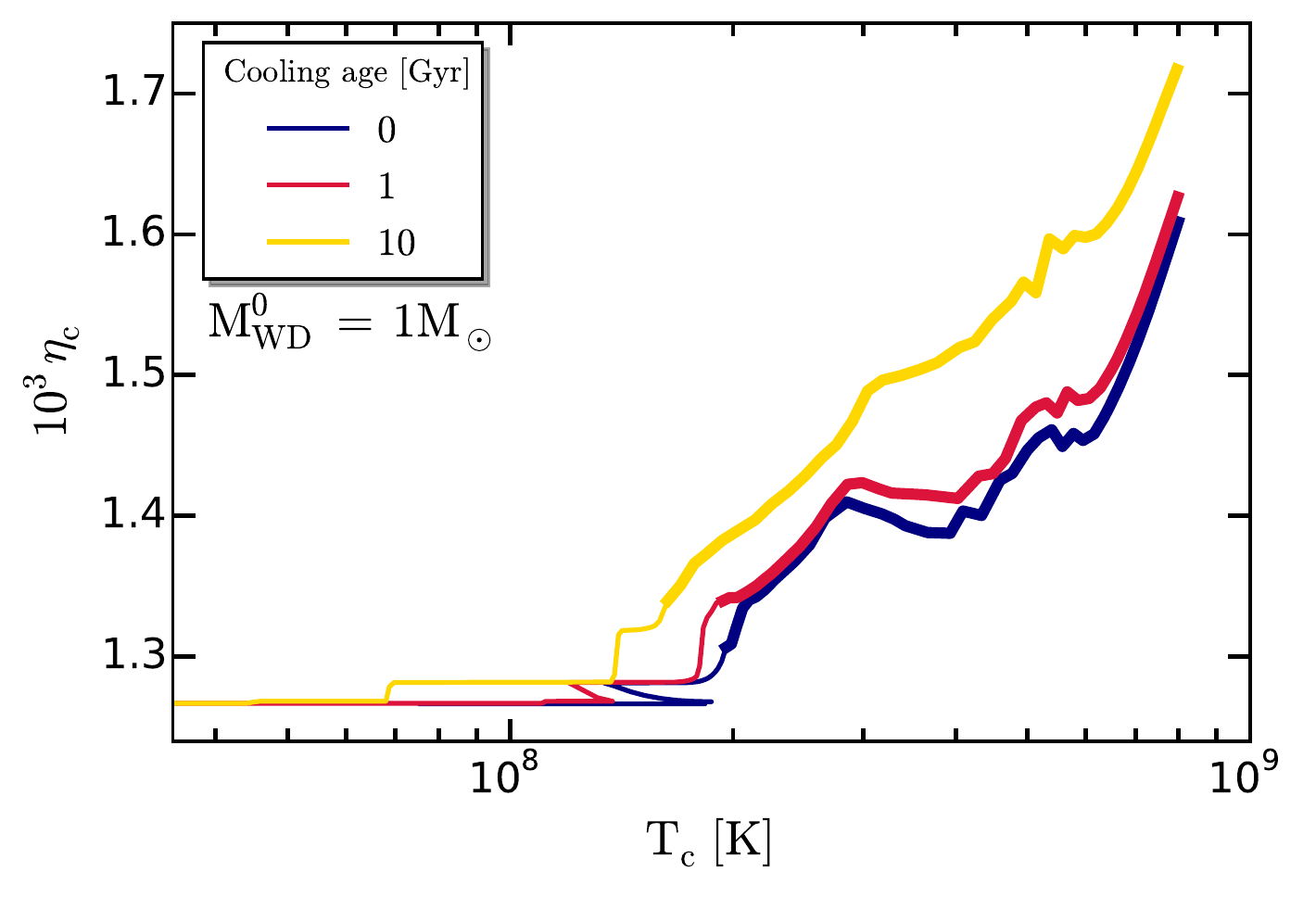}
\includegraphics[scale=0.61]{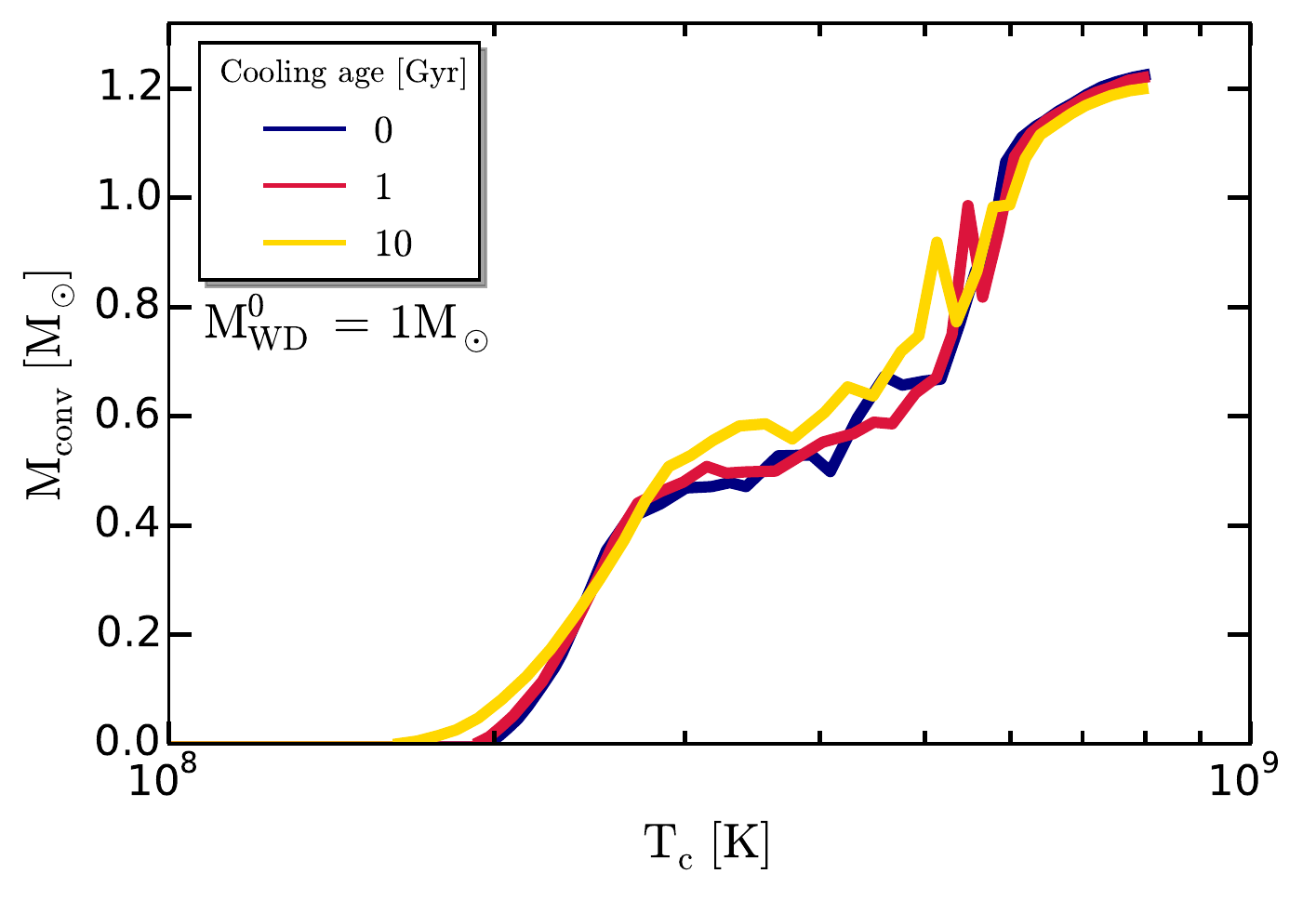}
\caption{The impact on the simmering of cooling ages equal to $0\,{\rm Gyr}$ (blue curve), $1\,{\rm Gyr}$ (red curve), and $10\,{\rm Gyr}$ (yellow curve). In each case, the simmering region is represented with thick lines. The top panel shows the evolution of the central temperature and the central density of a $1\,M_{\odot}$, solar-metallicity star with an accretion rate of $10^{-7} \, M_{\odot}\,{\rm yr^{-1}}$ and different cooling ages. The middle panel plots the evolution of the central neutron excess as a function of the central temperature. The bottom panel summarizes the growth of the mass of the convective core. Notice that the temperature limits are different in this plot.}
\label{etaTcMccRhoc_cooling}
\end{figure}

We next consider more broadly the results of the 135 models of our parameter survey. Figure \ref{etaTcMccRhoc_cooling} shows the behavior of the central density and temperature, the growth of the central neutron excess and the evolution of the convective core for the ``hot'' fiducial model discussed in Section \ref{Fiducial}, as well as the ``warm'' and ``cold'' versions of it. The effect of the Urca-process neutrino cooling disappears as the cooling age of the WD increases and the central temperature of the WD at the electron capture threshold density decreases. The local Urca-process cooling can also be appreciated in the evolutionary track of $\eta_{\mathrm{c}}$, where $T_{\mathrm{c}}$ decreases while $\eta_{\mathrm{c}}$ increases above its initial value $\eta_{\mathrm{c,0}}$ (which is mainly determined by the original abundance of $^{22}$Ne, as discussed in Section \ref{Simmering}). In addition, it decreases around $T_{\mathrm{c}} \approx 3 \times 10^{8} \, \rm{K}$ when the outer edge of the convection zone crosses the $^{23}$Na--$^{23}$Ne Urca shell.

The central neutron excess is slightly larger for the ``cold'' WD because the electron captures increase for higher densities. The central temperature at the onset of simmering is approximately the same for the three WDs, as well as the final extent of the convective core. At the onset of the thermal runaway, $\rho_{\rm{c}}$ is the main relic of the cooling process, whereas accretion has ``erased'' the memory of the initial mass of the WD.

\begin{figure}
\includegraphics[scale=0.62]{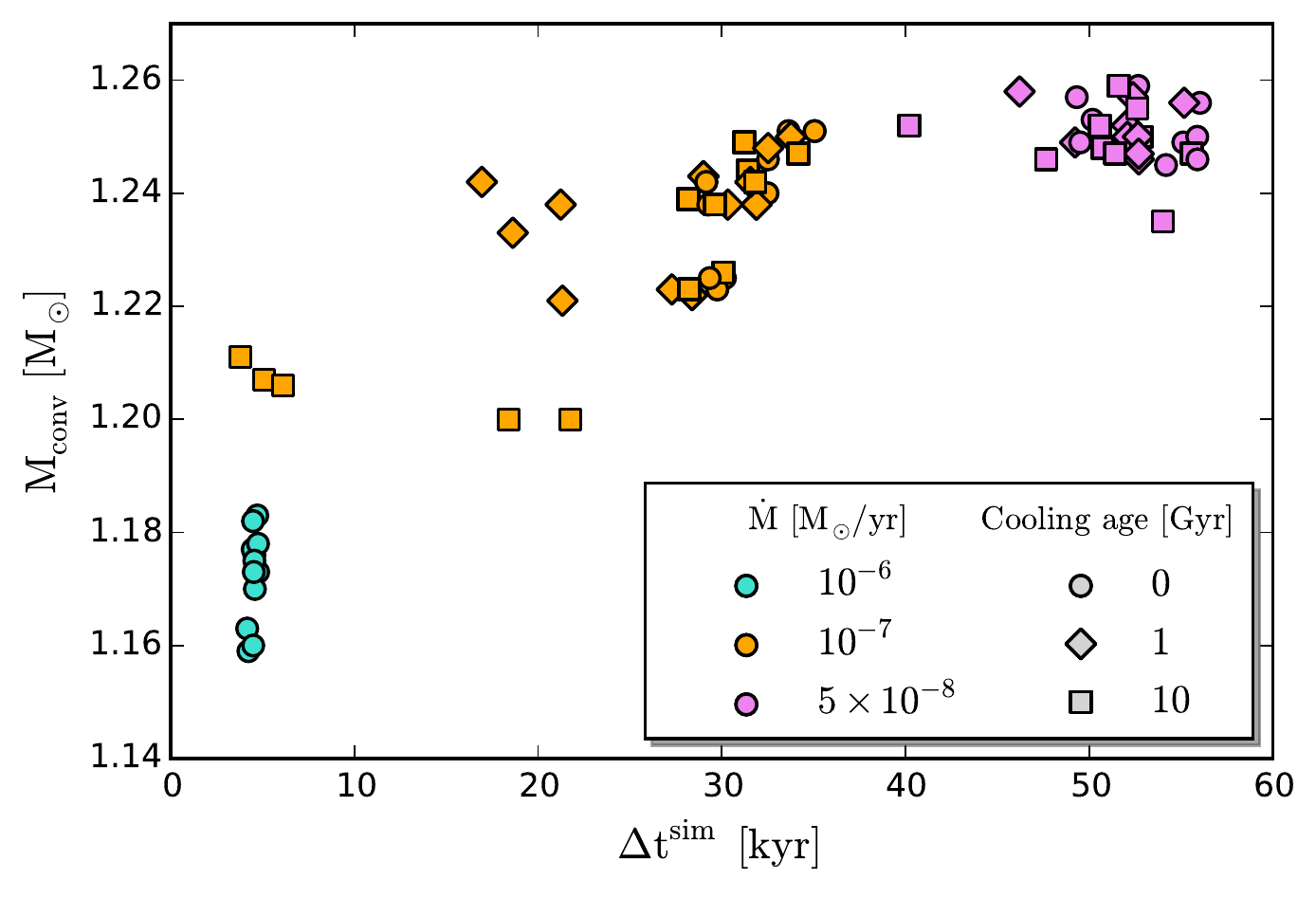}
\caption{Final mass of the convective core versus elapsed time during carbon simmering. Note that the different initial masses and metallicities are not labeled.}
\label{Panel1}
\end{figure}

\begin{figure}
\includegraphics[scale=0.62]{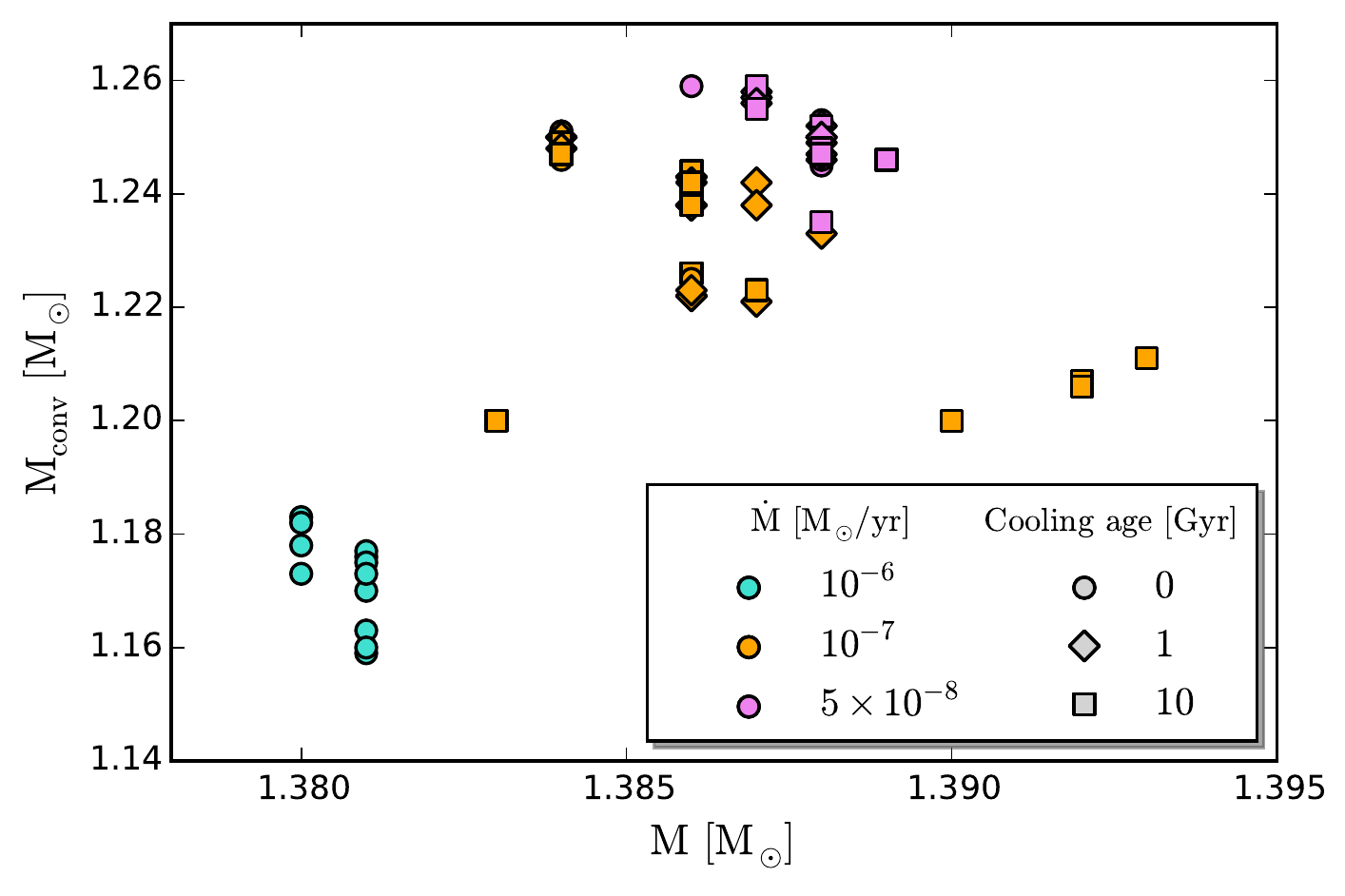}
\caption{Final mass of the convective core versus final mass.}
\label{Panel2}
\end{figure}

\begin{figure}
\includegraphics[scale=0.62]{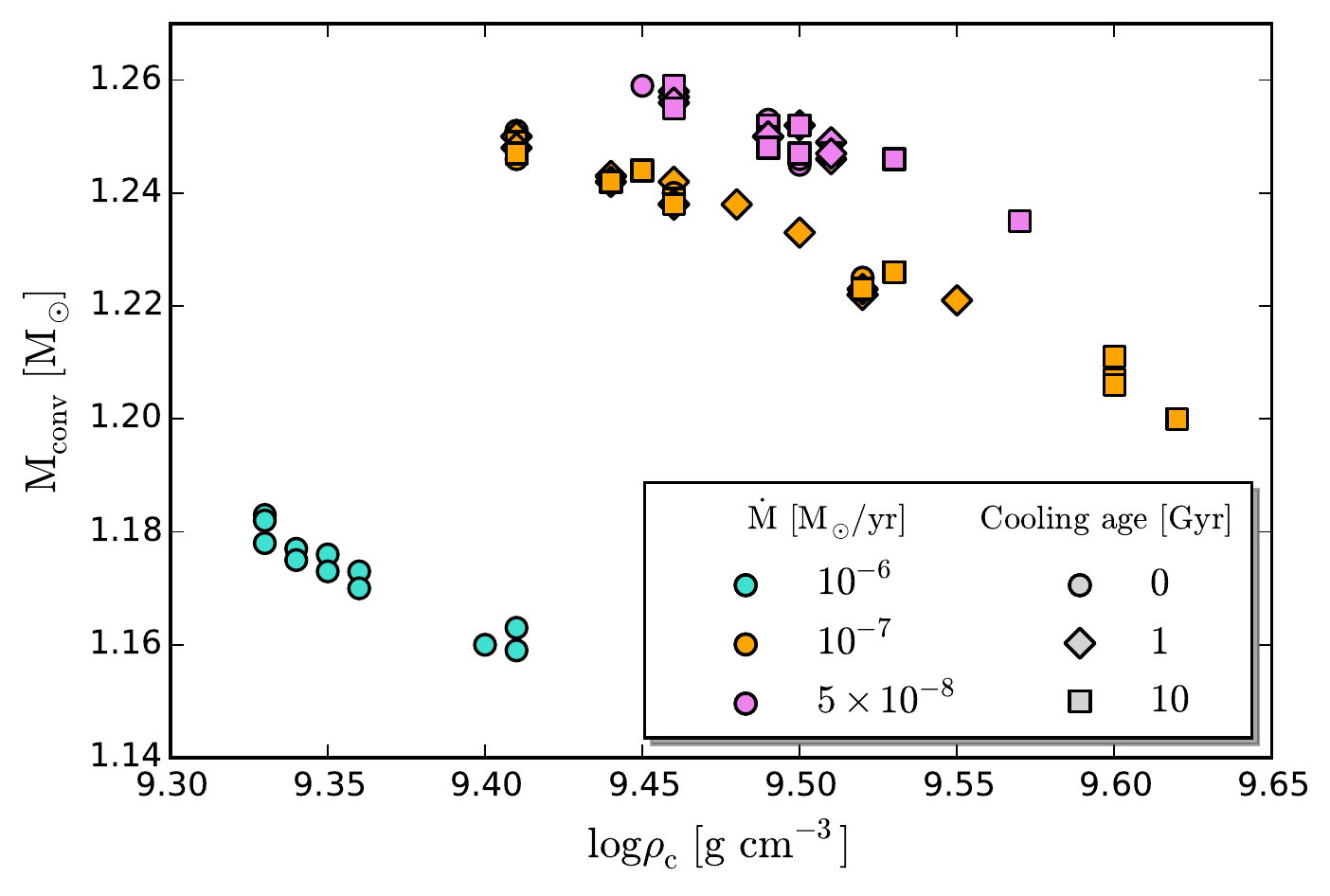}
\caption{Final mass of the convective core versus final central density.}
\label{Panel3}
\end{figure}

\begin{figure}
\includegraphics[scale=0.62]{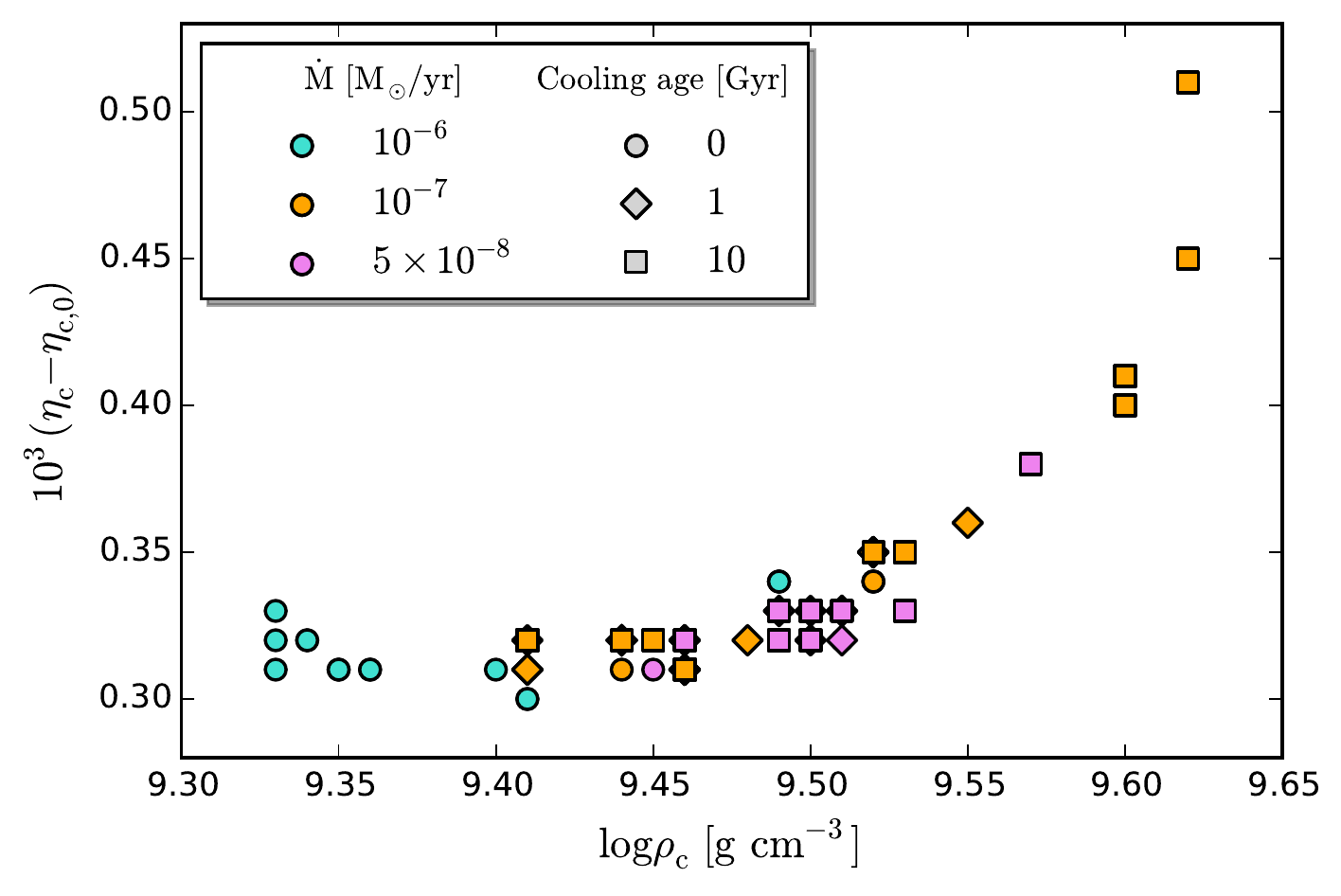}
\caption{Increase in the central neutron excess versus final central density.}
\label{Panel4}
\end{figure}

The remainder of our results are summarized in Figures \ref{Panel1}, \ref{Panel2}, \ref{Panel3}, and \ref{Panel4}, and in Tables \ref{Table_0Gyr}, \ref{Table_1Gyr} and \ref{Table_10Gyr}. Note that there are no fast accretors ($\dot{M} = 10^{-6} M_{\odot}\,{\rm yr^{-1}}$) in the case of the cooled WDs because they lead to off-center ignitions \citep{Ch14}.  In our tabulated results, we indicate these models with the note ``Off-center carbon ignition''.

\placefigure{f10}

\placefigure{f11}

Some of the general trends are as follows. The final masses of the convective core (shown in Figures \ref{Panel1}, \ref{Panel2}, and \ref{Panel3}) have relatively similar values $M_{\rm{conv}} \,\,{\approx}\,\, 1.16-1.26 \, M_{\odot}$, encompassing $\approx 85-90$\% of the final stars. This result agrees with the estimates of \citet{PiC08}. The elapsed times during simmering are longer when the convective core is larger (see Figure \ref{Panel1}) and are typically ${\gtrsim} \, 10^{4}$ years. The only models with $\Delta \rm{t} \, {\sim}  \,10^{3}$ years are the fast accretors and the ``cold'' WDs with an initial mass of $1 \, M_{\odot}$. Accretion times for these models are smaller and the shallower heat is unable to get to the core until a long time has elapsed. This, in turn, translates into higher ignition densities and more brief elapsed times during simmering. This is somewhat different from the estimate of $\Delta \rm{t} \sim 10^{3}$ years obtained by \citet{PiC08}, which was based on the central conditions. This work does note that a realistic value for the simmering time depends on the size of the region heated \citep[see Equation (8) of][which describes this]{PiC08}. The neutron excess increases with higher central densities (see Figure \ref{Panel4}) as the electron captures get more favored.

\begin{figure}
\centering
\includegraphics[scale=0.61]{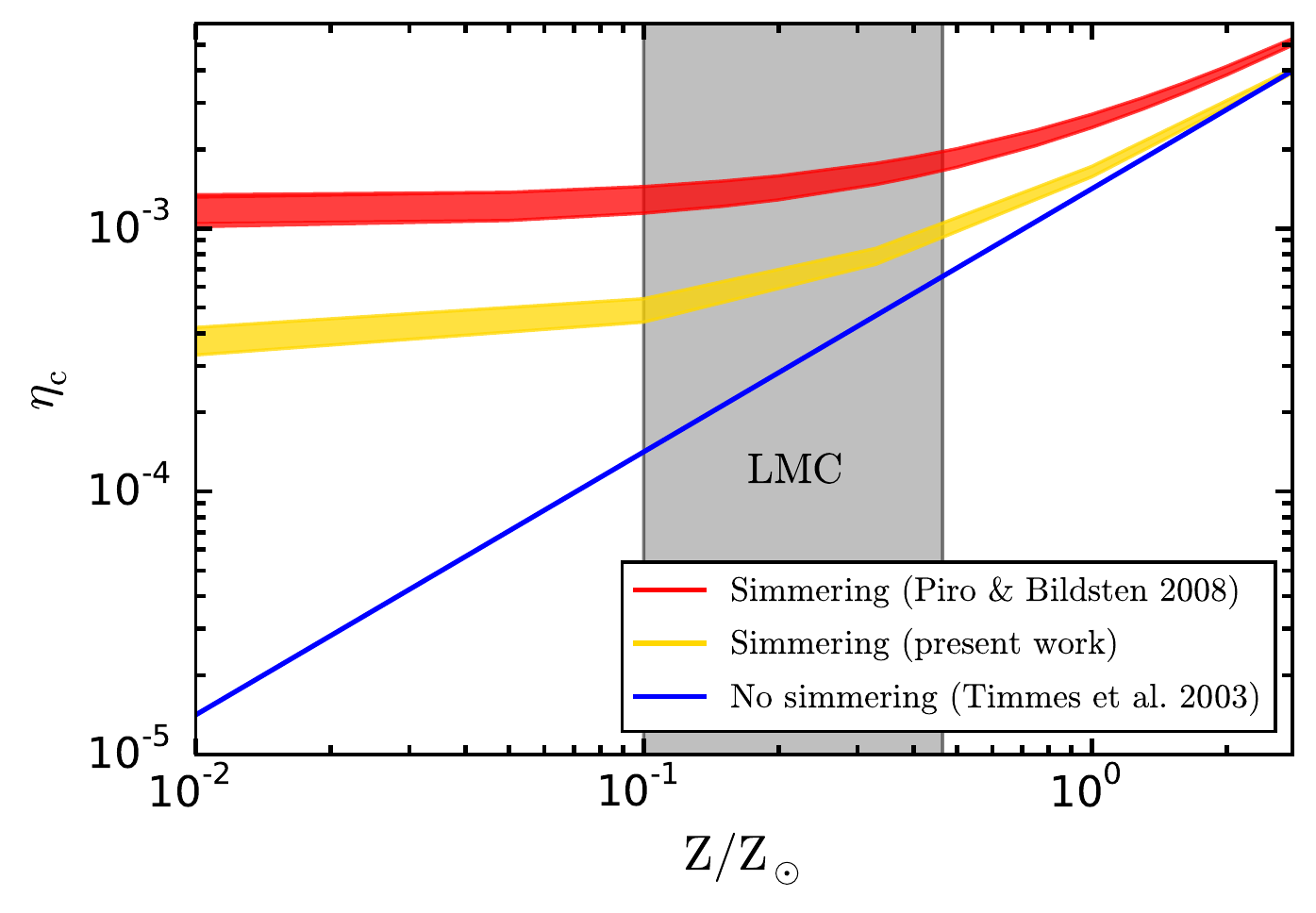}
\caption{The central neutron excess as a function of the metallicity of SNe Ia progenitors that experience no simmering (blue line), simmering according to \citet{PiB08} (red region), and simmering according to our work here (yellow region). This highlights the impact of the simmering floor at sufficiently low metallicities. Typical values of $Z$ for the Large Magellanic Cloud \citep{Pia13} are shown as a gray shaded region. Note that we use $Z_{\odot} = 0.014$ \citep{As09}.}
\label{etaZ_models}
\end{figure}

Finally, our results concerning the impact of simmering on the neutronization are summarized in Figure \ref{etaZ_models}, where we plot the expected neutron excess of a SN Ia progenitor versus its initial metallicity. The blue curve shows the linear relationship of $\eta = 0.101 Z$ derived by \citet{Ti03}. The red region shows the range of maximum neutronization estimates, in the range of 0.93$\, \eta_{\odot}$, from \citet{PiB08} and demonstrates the role played by the simmering floor. Namely, at sufficiently low metallicity, the neutron excess no longer reflects that of the progenitor but instead the amount of neutronization during simmering. Although we do find some small differences between models, Figure \ref{Panel4} demonstrates that the range of possible neutron excesses is relatively small, and thus we take the fiducial simmering limit to be 0.22$\, \eta_{\odot}$ (yellow, shaded region in Figure \ref{etaZ_models}), well below the value found by \citet{PiB08} for the reasons outlined in the discussions above.

\placefigure{f12}

\placefigure{f13}

\placefigure{f14}

\section{Conclusions} \label{Conclusions}

We have performed the first study of carbon simmering in SNe Ia progenitors with numerical models that fully resolve the extent of the convective region and include a complete nuclear network with Urca processes. We find that the final mass of the convective zone in the accreting WD is in the range of $M_{\rm conv} \,\,{\approx}\,\, 1.16-1.26\,M_\odot$. Our final values for the increase in the central neutron excess $\eta_{\mathrm{c}}$ before the onset of thermonuclear runaway are fairly constant at ${\approx}\,\, 3-4\times10^{-4}$. These values are ${\approx} \, 70 \%$ lower than those found by previous studies \citep{PiB08,Ch08}, with the difference stemming from our ability to properly resolve the full density profile of the convection zone and determine accurately where and at what rate electron captures occur. As the convection zone grows, it eventually spans many density scale heights, with electron captures favored in regions above the threshold density and beta decays favored in regions below it.  While the convective zone remains fully mixed, the overall neutronization is determined by the mass-weighted average of the reaction rates across the convection zone.

As summarized in Figure \ref{etaZ_models}, the lower simmering floor that we obtain makes it more challenging to find an observational ``smoking gun'' for the presence of simmering in SN Ia progenitors with metallicities $\gtrsim 1/3 \, Z_{\odot}$, typical of the thin disk of the Milky Way \citep{Nor04}. The strongest constraints on the degree of neutronization in individual SN Ia progenitors come from the analysis of the X-ray emission from Fe-peak nuclei (Mn, Cr, Fe, and Ni) in Galactic SNRs like Tycho, Kepler and 3C 397 \citep[see][]{Ba08,Park13,Ya15}. In the dynamically young SNRs Tycho and Kepler, where the bulk of the shocked Fe-peak elements were synthesized in the explosive Si burning regime \citep{Park13}, the Mn/Cr mass ratio is a clean tracer of progenitor neutronization. \citet{Ba08} and \cite{Park13} found a high level of neutronization in these SNRs, which translates to super-solar progenitor metallicities $Z/Z_{\odot} = 3.4^{+3.6}_{-2.6}$ and $Z/Z_{\odot} = 3.6^{+4.6}_{-2.0}$ if the contribution from simmering is neglected. The constraints on the progenitor neutronization in SNR 3C 397 are more model-dependent because this is a dynamically older object, and the shocked ejecta has a large contribution from neutron-rich NSE material. Nevertheless, \cite{Ya15} also found that, neglecting the contribution from simmering, Chandrasekhar-mass explosion models for this SNR require very high ($Z/Z_{\odot} \ {\sim} \ 5$) progenitor metallicities. These high levels of neutronization in Galactic Type Ia SNRs seem to be in tension with our results, because simmering is unable to reconcile the observations with a population of progenitors that is typical of the thin disk of the Milky Way, which contains very few stars with $Z/Z_{\odot} \gtrsim 3$. We hope to gain further insight on this apparent mismatch between models and observations by examining Type Ia SNRs in the LMC, which should have progenitor metallicities $\approx 0.1-0.4 Z_{\odot}$ \citep[][see Figure \ref{etaZ_models}]{Pia13}, low enough to clearly determine whether their progenitors underwent a carbon simmering phase and constrain the resulting degree of neutronization.

In the future, our models could be used as an input for SNe cosmology and explosion studies, as done by \citet{Mo15} and \citet{Pi15}, who created models with \texttt{MESA} and then employed a different code \citep{Mor15} to track the evolution of the supernova light curves. Using our models as inputs for explosive burning calculations would also be helpful for exploring the impact of the centrally neutron-enhanced core on the explosion and the resulting light curve \citep[e.g.][]{Br10}. For example, \citet{To09} and \citet{Ja10} studied the influence of $^{22}$Ne on the laminar flame speed, energy release, and nucleosynthesis during the SN explosion. The enhanced neutronization would have a similar impact, and although the influence of the $^{22}$Ne was found to be modest in these studies, we also predict spatial differences caused by the presence of the convection zone.

In addition, there are pieces of physics that could be added to our simmering models. As mentioned in Section~\ref{Models}, the gravitational settling of $^{22}\rm{Ne}$ will be implemented in an upcoming \texttt{MESA} release. We expect to revisit these models with a more complete approach including this process, as well as an in-depth treatment of the chemical diffusion and rotation during convection. \citet{PiC08} and \citet{Pi08} initially explored these effects with a series of semi-analytic models, and it will be interesting to revisit them with a more realistic treatment. The properties of the convective zone and neutron excess we found here are fairly homogeneous over a wide range of parameters. Therefore, it will be useful to see if other effects can add more diversity.

\acknowledgments
We thank the whole \texttt{MESA} community for their unconditional help during the elaboration of this paper, and especially Frank Timmes for useful discussions in regards to the implementation of nuclear reactions in \texttt{MESA}. We also thank Ed Brown and Remco Zegers for helpful communications regarding the ft-values for electron capture on $^{13}\rm{N}$, and Sumit Sarbadhicary for his initial work in the project. Finally, we are grateful to the anonymous referee, Dean Townsley and D. John Hillier for their useful feedback which helped improve the quality of this paper. This work has been funded by NASA ADAP grant NNX15AM03G S01. JS is supported by NSF grant AST-1205732.



\vspace{2 cm}
\appendix

\vspace{0.5 cm}
\section{Modifications to \texttt{MESA} and key weak reactions} \label{AppA}

In this appendix, we describe our use and extension of \texttt{MESA}'s on-the-fly weak rates capabilities.\footnote{\parbox{\textwidth}{We incorporate fixes that correct errors present in \citet{Pa15}, as documented in the published erratum (Paxton et al. 2016 in prep.).}} An accurate treatment of the key weak reaction rates is necessary to resolve the effects of the Urca process and to include the effects of neutronization due to the electron-capture reactions during simmering. In order to illustrate their importance, Figure~\ref{etavsTc_appendix} shows the differences between our work and a \texttt{MESA} calculation which does not include these choices and changes.

\vspace{0.5 cm}
\subsection{Weak rates for $A = 23$, $24$, and $25$}

\begin{figure}[H]
\epsscale{0.65}
\plotone{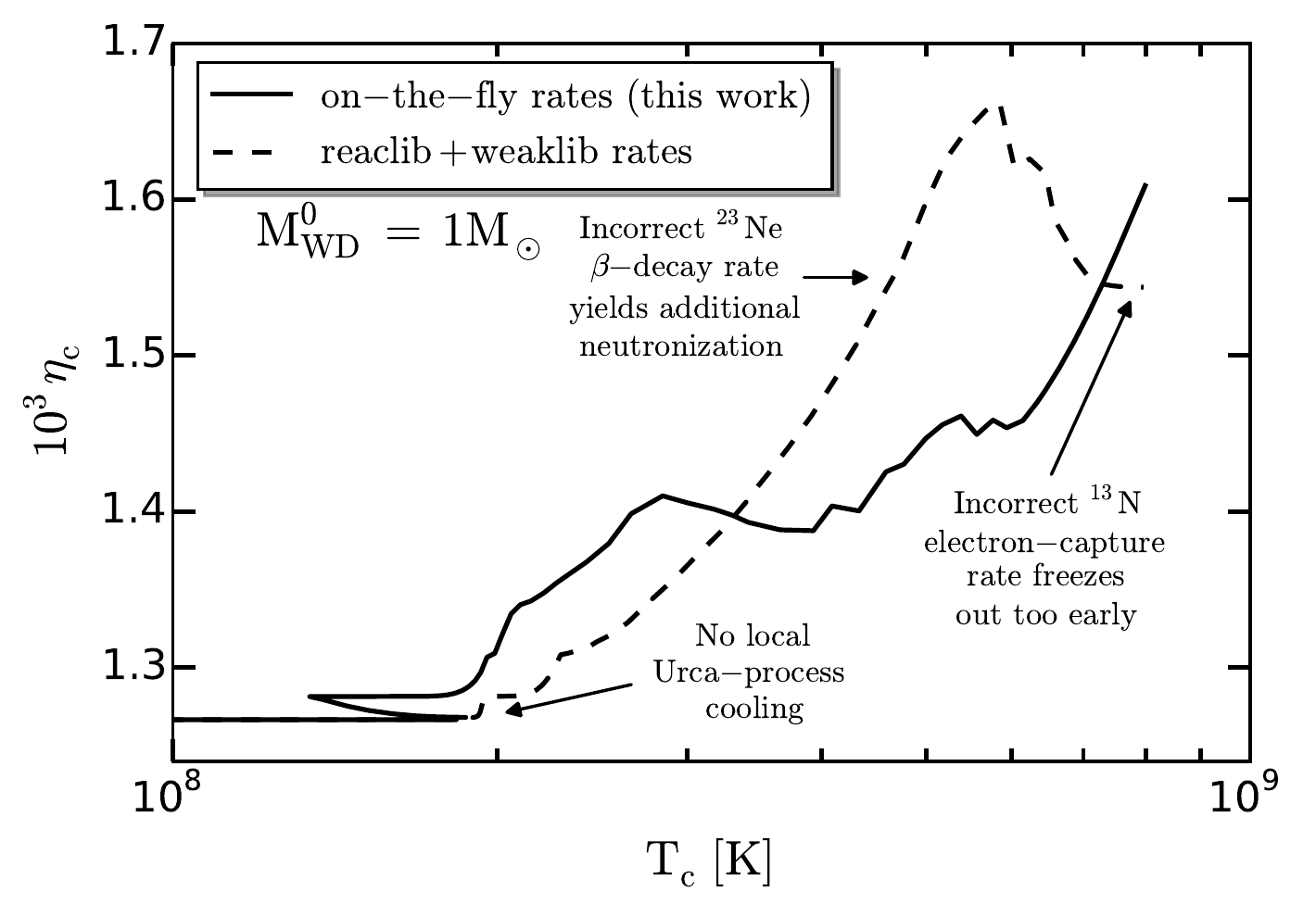}
\caption{Neutron excess as a function of central temperature for the fiducial model discussed Section \ref{Fiducial} with (black line) and without the use of the extended on-the-fly rates capabilities (dashed line).}
\label{etavsTc_appendix}
\end{figure}

Coarse tabulations of weak rates can severely underestimate cooling by the Urca process \citep[e.g.,][]{Toki13,Pa15}.  To circumvent this limitation, we use \texttt{MESA}'s capability to calculate weak reaction rates on-the-fly.  We use input nuclear data drawn from the \texttt{MESA} test suite problem \texttt{8.8M\_urca}, which includes Urca-process cooling by the $^{25}$Mg--$^{25}$Na and $^{23}$Na--$^{23}$Ne Urca pairs.  This choice allows us to include the significant and often neglected effects of local Urca process cooling via these isotopes (see Section \ref{Urca}). As indicated in Figure~\ref{etavsTc_appendix}, the decrease in temperature associated with the Urca-process cooling is not seen in a calculation that does not make use of the on-the-fly rates.
\\

The Urca-process cooling leads to an increase in the maximum central density reached (see Figure~\ref{TcvsRhoc_Urca}).  In some cases, this approaches or exceeds the threshold density for electron capture on $^{24}$Mg.  Therefore, we include weak reactions involving $^{24}$Mg and its daughters using input nuclear data drawn from the \texttt{MESA} test suite problem \texttt{wd\_aic}.
\\

The $\,^{23}$Na($\rm{e^{-}}$, $\rm{\nu_{e}}$)$^{23}$Ne$\,$ reaction plays a key role during the simmering phase (see Section \ref{Simmering}). As the convection zone grows, it eventually spans many density scale heights, with electron captures favored in regions above the threshold density and beta decays favored in regions below it.  While the convective zone remains fully mixed, the overall neutronization is determined by the mass-weighted average of the reaction rates across the convection zone.  Interpolation in the coarse \texttt{weaklib} tables leads to a systematic underestimate of the $^{23}$Ne beta-decay rates.  Figure~\ref{etavsTc_appendix} shows that a calculation using the on-the-fly rates exhibits less neutronization once the outer edge of the convection zone grows beyond the threshold density of $^{23}$Na.

\placefigure{f15}

\vspace{0.5 cm}
\subsection{Rate of electron capture on $\,^{13}$\textrm{N}}

In an unmodified version of \texttt{MESA} r7624, the reaction linking $^{13}$N to $^{13}$C (\texttt{r\_n13\_wk\_c13}) is drawn from JINA reaclib \citep{Cyburt10}.  This reaction rate includes only positron emission and does not include the electron-capture reaction $\,^{13}$N($\rm{e^{-}}$,$\rm{\nu_{e}}$)$^{13}$C.  At the characteristic simmering densities $(\rho \, {\sim} \, 10^9~\rm g\, cm^{-3})$, the electron capture rate is $\sim 10~\rm s^{-1}$, a factor of $\sim 10^{4}$ more rapid than the positron emission rate.  If the proper rate is not included, late in the simmering phase, $^{13}$N to $^{13}$C will freeze out.  This is illustrated on the right in Figure~\ref{etavsTc_appendix}, where the neutronization ceases to increase when \texttt{MESA}'s default \texttt{r\_n13\_wk\_c13} rate is used.
\\

The on-the-fly reaction rate framework described in \citet{Pa15} is limited to transitions with $Q<0$, where $Q$ is the energy difference (including rest mass) between the two states. The energy difference between the ground states of $\,^{13}$N and $^{13}$C is $Q = 2.22\,\mathrm{MeV}$ so, in order to incorporate this rate, we extend the on-the-fly weak rate implementation in \texttt{MESA} to include rates with $Q>0$.  In the notation of \citet{Pa15}, the rate for such a transition can be written as
\begin{equation}
  \lambda_{ij} = \frac{\ln 2}{(ft)_{ij}} \frac{\exp({\pi\alpha Z})}{(\me c^2)^5} \int_{\me c^2}^{\infty} \frac{E_{\rm e}^2 (E_{\rm e}+Q_{ij})^2}{1 + \exp[\beta (E_{\rm e} - \mue)]} dE_{\rm e}~.
\end{equation}
This integral can also be rewritten in terms of Fermi-Dirac integrals as in \citet{Sc15}, and as such, the extension is straightforward.  A patch demonstrating this implementation will be made available along with the inlists used in this work.
\\

We include the effects of two electron-capture transitions, drawing nuclear energy levels from \citet{AjzenbergSelove91} and $(ft)$-values from the recent experimental results of \citet{Zegers08}. These values are shown in Table \ref{tab:transitions}.

\begin{table}[H]
\caption{The transitions used in the on-the-fly $\,^{13}$N($\rm{e^{-}}$,$\rm{\nu_{e}}$)$^{13}$C rate calculation.  $E_\mathrm{i}$ and $E_\mathrm{f}$ are respectively the excitation energies (in MeV) of the initial and final states, relative to the ground state.  $J^\pi_\mathrm{i}$ and $J^\pi_\mathrm{f}$ are the spins and parities of the initial and final states. $(ft)$ is the comparative half-life in seconds.}
 \label{tab:transitions}
\centering
\begin{tabular}{ccccc}
\hline
$E_\mathrm{i}$ & $J_\mathrm{i}^\pi$ & $E_\mathrm{f}$  & $J_\mathrm{f}^{\pi}$ & $\log(ft)$ \\
\hline
0.000 ~&~ $1/2^{-}$ ~&~ 0.000 ~&~ $1/2^{-}$ ~&~ 3.665\\
0.000 ~&~ $1/2^{-}$ ~&~ 3.685 ~&~ $3/2^{-}$ ~&~ 3.460\\
\hline
\end{tabular}
\end{table}

\vspace{0.5 cm}
\section{Convergence of the models and overshooting} \label{AppB}

In this appendix, we address the numerical convergence of our models and the effects of overshooting. During the phase where the WD mass is in excess of $1.3 M_{\odot}$, which includes both the local Urca-process cooling
and simmering phases, the default spatial resolution of our models is specified by the control
\begin{verbatim}
    mesh_delta_coeff = 1.0    .
\end{verbatim}
The default temporal resolution of our models is specified by imposing a maximum allowed fractional change in the central density and temperature per timestep, via the controls
\begin{verbatim}
    delta_lgRho_cntr_hard_limit = 1d-3
    delta_lgT_cntr_hard_limit = 3d-3    .
\end{verbatim}

\begin{figure}[H]
\epsscale{0.55}
\plotone{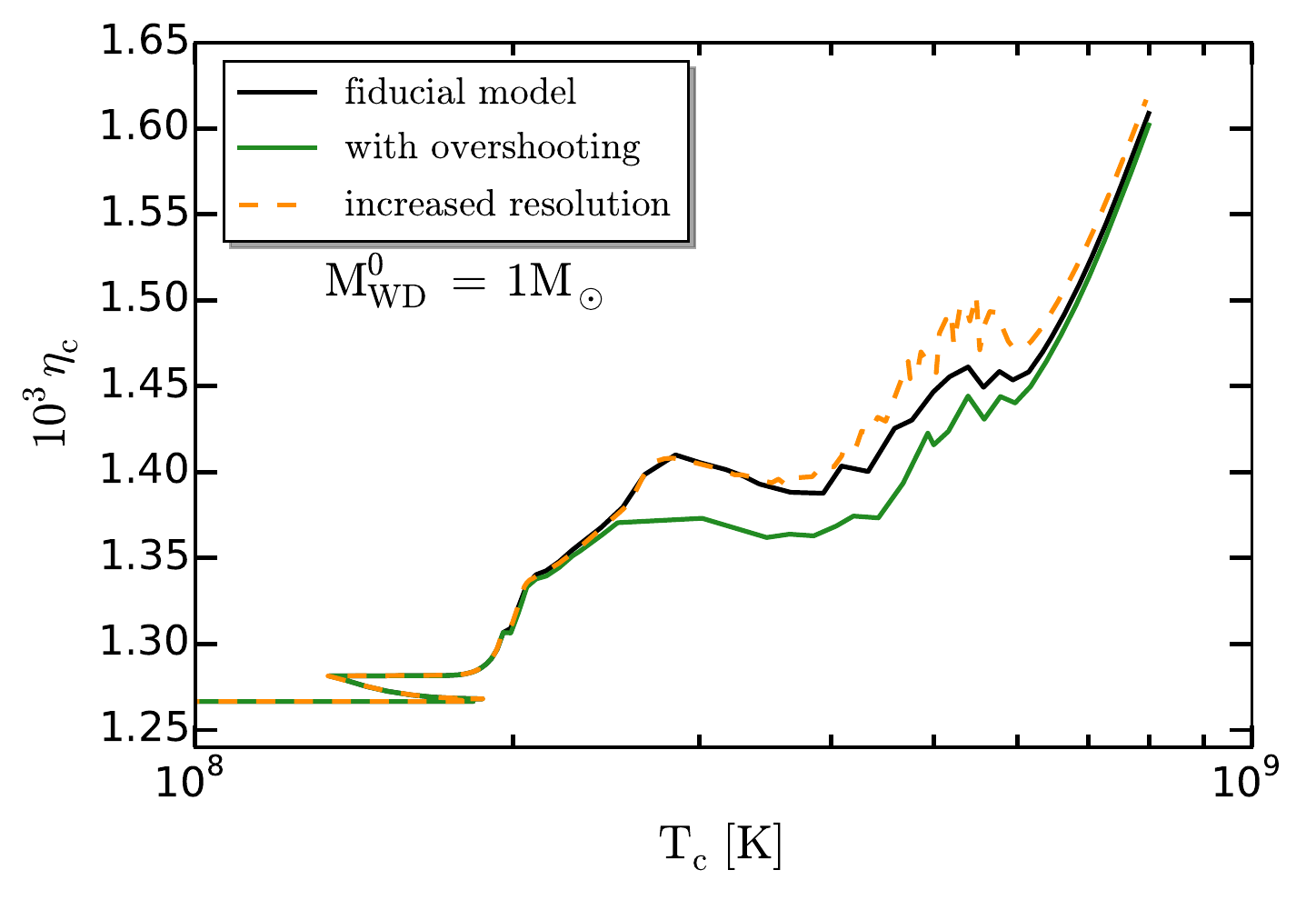}
\epsscale{0.55}
\plotone{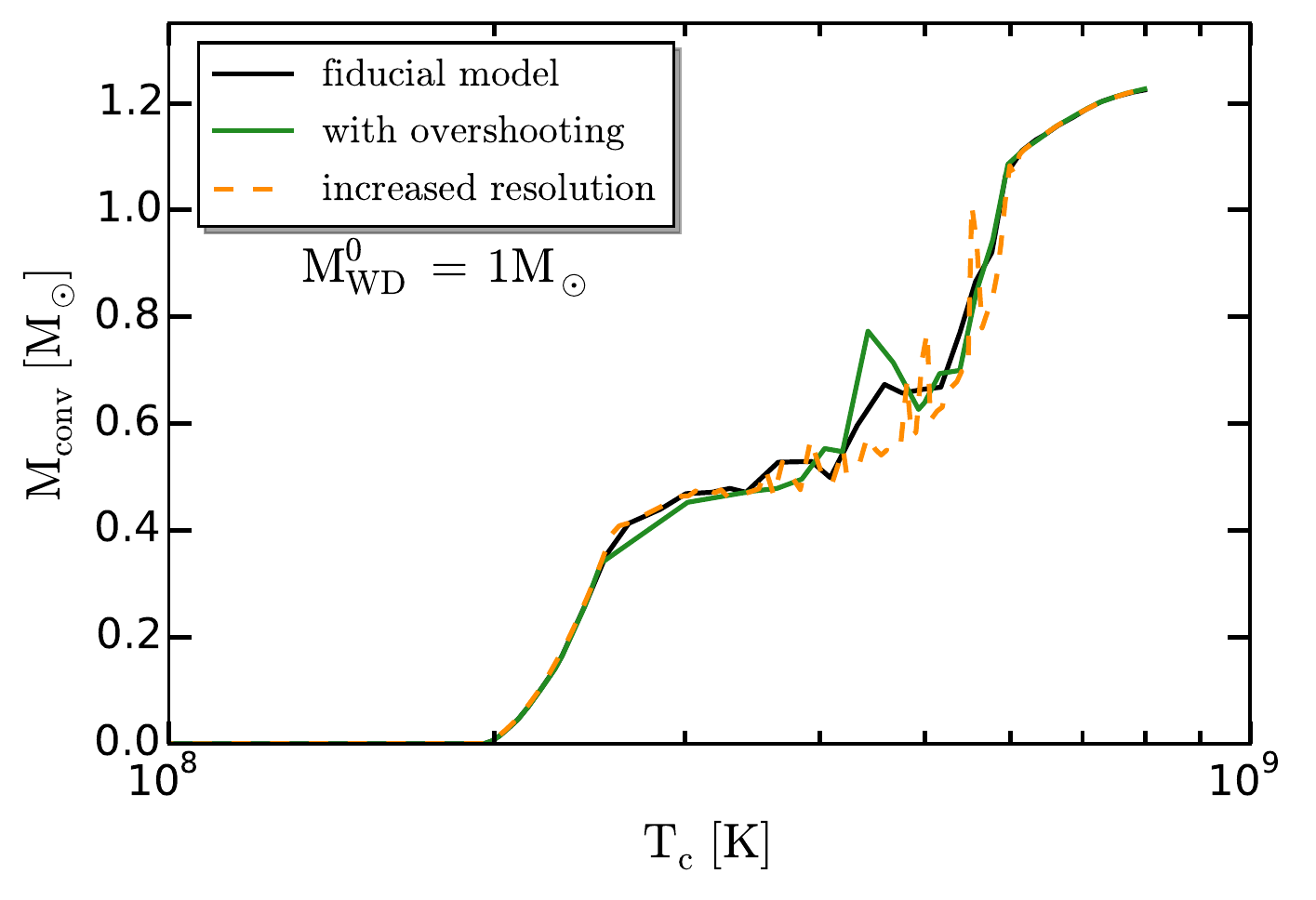}
\caption{Comparison of results from the fiducial model (black curve), a model with overshooting (green curve) and a model with increased spatial and temporal resolution (dashed, orange curve). Left: Central neutron excess. Right: Mass of the convective core.  The primary differences occur for $T_\mathrm{c}$ between $3\times 10^8$ K and $6\times 10^8$ K.  During this phase, our limited treatment of the convective Urca-process makes fine details of the models unlikely to be physically meaningful.  By the end of the evolution, the models return to a smooth evolution and to good agreement with each other. }
\label{etaMccvsTc_mesh015hl}
\end{figure}

In order to confirm that our results are robust, we repeated our fiducial calculation, but used these controls to increase the spatial resolution by a factor of $\approx 6$ and the temporal resolution by a factor of $\approx 3$. Figure \ref{etaMccvsTc_mesh015hl} compares our fiducial model (Section \ref{Fiducial}) to this model with increased temporal and spatial resolution.  The most conspicuous differences occur for $T_\mathrm{c}$ between $3\times 10^8$ K and $6\times 10^8$ K.  It is primarily during this phase that the convective Urca-process is occurring.  As mentioned in Section \ref{Urca}, limitations imposed by the 1D mixing length theory of convection prevent fully consistent modeling of this phase. Thus, the fine details of the observed behavior during this phase are unlikely to be physically meaningful.  As simmering nears its end and the heating timescale falls, the Urca-process reactions begin to freeze out.  Once this occurs, the models return to a smooth evolution and to good agreement with each other. 
\\

The models shown in the body of the paper use the Schwarzschild convective criterion and no overshooting.  Figure \ref{etaMccvsTc_mesh015hl} also shows the results of our fiducial model with overshooting at the outer boundary of the central convective zone, added by means of the controls
\begin{verbatim}
    overshoot_f_above_burn_z_core = 0.010
    overshoot_f0_above_burn_z_core = 0.005  .
\end{verbatim}
Again, the primary differences occur for $T_\mathrm{c}$ between $3\times 10^8$ K and $6\times 10^8$ K, but the model returns to good agreement with the fiducial model by the end of simmering.
In Table~\ref{Table_convergence} we compare the values of the quantities of interest at the end of simmering (the same quantities compiled in Tables~\ref{Table_0Gyr}--\ref{Table_10Gyr}) for our fiducial model, the model with increased resolution, and the model including the effects of overshooting.  We find sub-to-few per cent level agreement in all quantities of interest, giving us confidence that our results are robust.

\begin{table}[H]
\parbox{\textwidth}{\caption{Comparison of results from the fiducial model (top), a different run with overshooting (middle) and a model with increased spatial and temporal resolution (bottom). The number of cells at the end of the run and the total number of time steps are shown in the last two columns.} \label{Table_convergence}} 
\centering
\begin{tabular}{lccccccccccc}
\tableline\tableline
\noalign{\smallskip}
Description & $\log T\rm{_{c}^{\, sim}}$ & $\rm{\log \rho_{c}^{\, sim}}$ & $\rm{\Delta t^{sim}}$ & $\rm{t}$ & $\rm{\log \rho_{c}}$ & $M_{\rm{WD}}$ & $M_{\rm{conv}}$ & $\rm{10^{3} \eta_{c}}$ & $\rm{10^{3} (\eta_{c}-\eta_{c,0}})$ & \# Zones & Time steps \\
\noalign{\smallskip}
& $[\rm{K}]$ & $[\rm{g\,cm^{-3}}]$ & $[\rm{kyr}]$ & $[\rm{Myr}]$ & $[\rm{g\,cm^{-3}}]$ & $[M_{\odot}]$ & $[M_{\odot}]$ & $\mathrm{}$ & & & \\
\noalign{\smallskip}
\tableline
\noalign{\smallskip}
Fiducial model & 8.29 & 9.61 & 30.2 & 3.86 & 9.52 & 1.386 & 1.225 & 1.61 & 0.34 & 1149 & 1677\\ 
With overshooting & 8.29 & 9.61 & 29.6 & 3.86 & 9.52 & 1.386 & 1.227 & 1.60 & 0.34 & 1144 & 1640\\ 
Increased resolution & 8.29 & 9.61 & 31.0 & 3.86 & 9.53 & 1.386 & 1.225 & 1.62 & 0.35 & 7235 & 5343\\ 
\noalign{\smallskip}
\tableline
\end{tabular}
\end{table}

\placefigure{f16a}
\placefigure{f16b}




\vspace{0.5 cm}


\twocolumngrid


\begin{table}		
\vspace{0.1 cm} 
\parbox{\textwidth}{\caption{Results for the models without cooling.} \label{Table_0Gyr}} 
\centering
\noindent\makebox[\textwidth]{ 
\begin{tabular}{lccccccccccc}
\tableline\tableline
\noalign{\smallskip}
$Z$ & $M_{\rm{WD}}^{0}$ & $\dot{M}$ & $\log T\rm{_{c}^{\, sim}}$ & $\rm{\log \rho_{c}^{\, sim}}$ & $\rm{\Delta t^{sim}}$ & $\rm{t}$ & $\rm{\log \rho_{c}}$ & $M_{\rm{WD}}$ & $M_{\rm{conv}}$ & $\rm{10^{3} \eta_{c}}$ & $\rm{10^{3} (\eta_{c}-\eta_{c,0}})$ \\
\noalign{\smallskip}
$[Z_{\odot}]$ & $[M_{\odot}]$ &  $[M_{\odot}\,\rm{yr}^{-1}]$ & $[\rm{K}]$ & $[\rm{g\,cm^{-3}}]$ & [$\rm{kyr}]$ & $[\rm{Myr}]$ & $[\rm{g\,cm^{-3}}]$ & $[M_{\odot}]$ & $[M_{\odot}]$ & $\mathrm{}$ & $\mathrm{}$ \\
\noalign{\smallskip}
\tableline
\noalign{\smallskip}
0.01 & 0.7 & $10^{-6}$ & 8.43 & 9.38 & 4.76 & 0.680 & 9.33 & 1.380 & 1.178 & 0.34 & 0.33 \\
0.01 & 0.85 & $10^{-6}$ & 8.43 & 9.38 & 4.49 & 0.529 & 9.33 & 1.380 & 1.182 & 0.34 & 0.32 \\
0.01 & 1.0 & $10^{-6}$ & 8.43 & 9.38 & 4.72 & 0.379 & 9.33 & 1.380 & 1.183 & 0.33 & 0.31 \\
0.01 & 0.7 & $10^{-7}$ & 8.36 & 9.50 & 32.5 & 6.84 & 9.41 & 1.384 & 1.246 & 0.33 & 0.32 \\
0.01 & 0.85 & $10^{-7}$ & 8.35 & 9.49 & 35.1 & 5.34 & 9.41 & 1.384 & 1.251 & 0.33 & 0.32 \\
0.01 & 1.0 & $10^{-7}$ & 8.36 & 9.50 & 33.6 & 3.84 & 9.41 & 1.384 & 1.251 & 0.33 & 0.31 \\
0.01 & 0.7 & $5 \times 10^{-8}$ & 8.32 & 9.56 & 56.0 & 13.7 & 9.46 & 1.387 & 1.256 & 0.34 & 0.32 \\
0.01 & 0.85 & $5 \times 10^{-8}$ & 8.33 & 9.56 & 49.3 & 10.7 & 9.46 & 1.387 & 1.257 & 0.33 & 0.32 \\
0.01 & 1.0 & $5 \times 10^{-8}$ & 8.32 & 9.56 & 52.7 & 7.73 & 9.45 & 1.386 & 1.259 & 0.33 & 0.31 \\
0.10 & 0.7 & $10^{-6}$ & 8.42 & 9.39 & 4.56 & 0.681 & 9.34 & 1.381 & 1.175 & 0.45 & 0.32 \\
0.10 & 0.85 & $10^{-6}$ & 8.42 & 9.39 & 4.47 & 0.531 & 9.34 & 1.381 & 1.177 & 0.45 & 0.32 \\
0.10 & 1.0 & $10^{-6}$ & 8.42 & 9.39 & 4.56 & 0.380 & 9.35 & 1.381 & 1.176 & 0.44 & 0.31 \\
0.10 & 0.7 & $10^{-7}$ & 8.34 & 9.54 & 29.2 & 6.86 & 9.44 & 1.386 & 1.242 & 0.45 & 0.32 \\
0.10 & 0.85 & $10^{-7}$ & 8.34 & 9.54 & 29.2 & 5.35 & 9.44 & 1.386 & 1.242 & 0.45 & 0.32 \\
0.10 & 1.0 & $10^{-7}$ & 8.33 & 9.53 & 31.5 & 3.85 & 9.44 & 1.386 & 1.242 & 0.44 & 0.31 \\
0.10 & 0.7 & $5 \times 10^{-8}$ & 8.30 & 9.60 & 49.5 & 13.8 & 9.49 & 1.388 & 1.249 & 0.46 & 0.33 \\
0.10 & 0.85 & $5 \times 10^{-8}$ & 8.30 & 9.59 & 55.9 & 10.8 & 9.49 & 1.388 & 1.250 & 0.46 & 0.33 \\
0.10 & 1.0 & $5 \times 10^{-8}$ & 8.30 & 9.60 & 50.2 & 7.75 & 9.49 & 1.388 & 1.253 & 0.46 & 0.33 \\
0.33 & 0.7 & $10^{-6}$ & 8.41 & 9.40 & 4.53 & 0.681 & 9.35 & 1.381 & 1.173 & 0.73 & 0.31 \\
0.33 & 0.85 & $10^{-6}$ & 8.41 & 9.40 & 4.59 & 0.531 & 9.36 & 1.381 & 1.170 & 0.73 & 0.31 \\
0.33 & 1.0 & $10^{-6}$ & 8.41 & 9.40 & 4.77 & 0.380 & 9.36 & 1.380 & 1.173 & 0.73 & 0.31 \\
0.33 & 0.7 & $10^{-7}$ & 8.33 & 9.55 & 29.3 & 6.86 & 9.46 & 1.386 & 1.238 & 0.73 & 0.31 \\
0.33 & 0.85 & $10^{-7}$ & 8.32 & 9.55 & 32.5 & 5.36 & 9.46 & 1.386 & 1.240 & 0.73 & 0.31 \\
0.33 & 1.0 & $10^{-7}$ & 8.32 & 9.55 & 31.8 & 3.85 & 9.46 & 1.386 & 1.242 & 0.73 & 0.31 \\
0.33 & 0.7 & $5 \times 10^{-8}$ & 8.29 & 9.61 & 55.9 & 13.8 & 9.50 & 1.388 & 1.246 & 0.74 & 0.32 \\
0.33 & 0.85 & $5 \times 10^{-8}$ & 8.29 & 9.61 & 54.2 & 10.8 & 9.50 & 1.388 & 1.245 & 0.75 & 0.32 \\
0.33 & 1.0 & $5 \times 10^{-8}$ & 8.29 & 9.61 & 55.1 & 7.75 & 9.51 & 1.388 & 1.249 & 0.74 & 0.32 \\
1.00 & 0.7 & $10^{-6}$ & 8.39 & 9.45 & 4.50 & 0.681 & 9.40 & 1.381 & 1.160 & 1.58 & 0.31 \\
1.00 & 0.85 & $10^{-6}$ & 8.40 & 9.46 & 4.24 & 0.531 & 9.41 & 1.381 & 1.159 & 1.57 & 0.30 \\
1.00 & 1.0 & $10^{-6}$ & 8.40 & 9.46 & 4.17 & 0.381 & 9.41 & 1.381 & 1.163 & 1.57 & 0.30 \\
1.00 & 0.7 & $10^{-7}$ & 8.29 & 9.61 & 29.4 & 6.86 & 9.52 & 1.386 & 1.225 & 1.61 & 0.35 \\
1.00 & 0.85 & $10^{-7}$ & 8.29 & 9.61 & 29.8 & 5.36 & 9.52 & 1.386 & 1.223 & 1.61 & 0.34 \\
1.00 & 1.0 & $10^{-7}$ & 8.29 & 9.61 & 30.2 & 3.86 & 9.52 & 1.386 & 1.225 & 1.61 & 0.34 \\
1.00 & 0.7 & $5 \times 10^{-8}$ & \multicolumn{9}{c}{Convection zone splits during simmering} \\
1.00 & 0.85 & $5 \times 10^{-8}$ & \multicolumn{9}{c}{Convection zone splits during simmering} \\
1.00 & 1.0 & $5 \times 10^{-8}$ & \multicolumn{9}{c}{Convection zone splits during simmering} \\
2.79 & 0.7 & $10^{-6}$ & 8.37 & 9.55 & 3.33 & 0.679 & 9.49 & 1.379 & 1.131 & 3.86 & 0.34 \\
2.79 & 0.85 & $10^{-6}$ & 8.37 & 9.55 & 3.64 & 0.529 & 9.49 & 1.379 & 1.136 & 3.86 & 0.34 \\
2.79 & 1.0 & $10^{-6}$ & 8.37 & 9.55 & 3.62 & 0.378 & 9.49 & 1.378 & 1.136 & 3.86 & 0.34 \\
2.79 & 0.7 & $10^{-7}$ & \multicolumn{9}{c}{Convection zone splits during simmering} \\
2.79 & 0.85 & $10^{-7}$ & \multicolumn{9}{c}{Convection zone splits during simmering} \\
2.79 & 1.0 & $10^{-7}$ & \multicolumn{9}{c}{Convection zone splits during simmering} \\
2.79 & 0.7 & $5 \times 10^{-8}$ & \multicolumn{9}{c}{Convection zone splits during simmering} \\
2.79 & 0.85 & $5 \times 10^{-8}$ & \multicolumn{9}{c}{Convection zone splits during simmering} \\
2.79 & 1.0 & $5 \times 10^{-8}$ & \multicolumn{9}{c}{Convection zone splits during simmering} \\
\noalign{\smallskip}
\tableline
\end{tabular}
}
\end{table}		
\clearpage

\begin{table}
\parbox{\textwidth}{{\caption{Results for the models with a cooling age of 1 Gyr.}\label{Table_1Gyr}}}
\centering
\noindent\makebox[\textwidth]{ 
\begin{tabular}{lccccccccccc}
\tableline\tableline
\noalign{\smallskip}
$Z$ & $M_{\rm{WD}}^{0}$ & $\dot{M}$ & $\log T\rm{_{c}^{\, sim}}$ & $\rm{\log \rho_{c}^{\, sim}}$ & $\rm{\Delta t^{sim}}$ & $\rm{t}$ & $\rm{\log \rho_{c}}$ & $M_{\rm{WD}}$ & $M_{\rm{conv}}$ & $\rm{10^{3} \eta_{c}}$ & $\rm{10^{3} (\eta_{c}-\eta_{c,0}})$ \\
\noalign{\smallskip}
$[Z_{\odot}]$ & $[M_{\odot}]$ &  $[M_{\odot}\,\rm{yr}^{-1}]$ & $[\rm{K}]$ & $[\rm{g\,cm^{-3}}]$ & [$\rm{kyr}]$ & $[\rm{Myr}]$ & $[\rm{g\,cm^{-3}}]$ & $[M_{\odot}]$ & $[M_{\odot}]$ & $\mathrm{}$ & $\mathrm{}$ \\
\noalign{\smallskip}
\tableline
\noalign{\smallskip}
0.01 & 0.7 & $10^{-6}$ & \multicolumn{9}{c}{Off-center carbon ignition} \\
0.01& 0.85 & $10^{-6}$ & \multicolumn{9}{c}{Off-center carbon ignition} \\
0.01& 1.0 & $10^{-6}$ & \multicolumn{9}{c}{Off-center carbon ignition} \\
0.01 & 0.7 & $10^{-7}$ & 8.36 & 9.50 & 32.5 & 6.84 & 9.41 & 1.384 & 1.248 & 0.33 & 0.32 \\
0.01 & 0.85 & $10^{-7}$ & 8.36 & 9.50 & 33.8 & 5.34 & 9.41 & 1.384 & 1.250 & 0.33 & 0.31 \\
0.01 & 1.0 & $10^{-7}$ & 8.35 & 9.57 & 17.0 & 3.87 & 9.46 & 1.387 & 1.242 & 0.33 & 0.31 \\
0.01 & 0.7 & $5 \times 10^{-8}$ & 8.32 & 9.56 & 55.2 & 13.7 & 9.46 & 1.387 & 1.256 & 0.34 & 0.32 \\
0.01 & 0.85 & $5 \times 10^{-8}$ & 8.32 & 9.56 & 52.4 & 10.7 & 9.46 & 1.387 & 1.257 & 0.33 & 0.32 \\
0.01 & 1.0 & $5 \times 10^{-8}$ & 8.32 & 9.57 & 46.2 & 7.72 & 9.46 & 1.387 & 1.258 & 0.33 & 0.31 \\
0.10& 0.7 & $10^{-6}$ & \multicolumn{9}{c}{Off-center carbon ignition} \\
0.10& 0.85 & $10^{-6}$ & \multicolumn{9}{c}{Off-center carbon ignition} \\
0.10& 1.0 & $10^{-6}$ & \multicolumn{9}{c}{Off-center carbon ignition} \\
0.10 & 0.7 & $10^{-7}$ & 8.33 & 9.53 & 31.6 & 6.85 & 9.44 & 1.386 & 1.242 & 0.45 & 0.32 \\
0.10 & 0.85 & $10^{-7}$ & 8.34 & 9.54 & 29.0 & 5.36 & 9.44 & 1.386 & 1.243 & 0.45 & 0.32 \\
0.10 & 1.0 & $10^{-7}$ & 8.32 & 9.58 & 21.2 & 3.87 & 9.48 & 1.387 & 1.238 & 0.45 & 0.32 \\
0.10 & 0.7 & $5 \times 10^{-8}$ & 8.30 & 9.59 & 52.7 & 13.8 & 9.49 & 1.388 & 1.250 & 0.46 & 0.33 \\
0.10 & 0.85 & $5 \times 10^{-8}$ & 8.30 & 9.59 & 52.1 & 10.8 & 9.49 & 1.388 & 1.250 & 0.46 & 0.33 \\
0.10 & 1.0 & $5 \times 10^{-8}$ & 8.29 & 9.60 & 52.1 & 7.76 & 9.50 & 1.388 & 1.252 & 0.46 & 0.33 \\
0.33& 0.7 & $10^{-6}$ & \multicolumn{9}{c}{Off-center carbon ignition} \\
0.33& 0.85 & $10^{-6}$ & \multicolumn{9}{c}{Off-center carbon ignition} \\
0.33& 1.0 & $10^{-6}$ & \multicolumn{9}{c}{Off-center carbon ignition} \\
0.33 & 0.7 & $10^{-7}$ & 8.33 & 9.55 & 30.3 & 6.86 & 9.46 & 1.386 & 1.238 & 0.73 & 0.31 \\
0.33 & 0.85 & $10^{-7}$ & 8.32 & 9.55 & 31.9 & 5.35 & 9.46 & 1.386 & 1.238 & 0.73 & 0.31 \\
0.33 & 1.0 & $10^{-7}$ & 8.32 & 9.61 & 18.6 & 3.87 & 9.50 & 1.388 & 1.233 & 0.74 & 0.32 \\
0.33 & 0.7 & $5 \times 10^{-8}$ & 8.29 & 9.61 & 52.7 & 13.8 & 9.51 & 1.388 & 1.247 & 0.75 & 0.32 \\
0.33 & 0.85 & $5 \times 10^{-8}$ & 8.29 & 9.61 & 49.2 & 10.8 & 9.51 & 1.388 & 1.249 & 0.75 & 0.33 \\
0.33 & 1.0 & $5 \times 10^{-8}$ & 8.28 & 9.61 & 52.7 & 7.76 & 9.51 & 1.388 & 1.246 & 0.75 & 0.33 \\
1.00& 0.7 & $10^{-6}$ & \multicolumn{9}{c}{Off-center carbon ignition} \\
1.00& 0.85 & $10^{-6}$ & \multicolumn{9}{c}{Off-center carbon ignition} \\
1.00& 1.0 & $10^{-6}$ & \multicolumn{9}{c}{Off-center carbon ignition} \\
1.00 & 0.7 & $10^{-7}$ & 8.30 & 9.62 & 27.3 & 6.86 & 9.52 & 1.386 & 1.223 & 1.62 & 0.35 \\
1.00 & 0.85 & $10^{-7}$ & 8.29 & 9.62 & 28.4 & 5.36 & 9.52 & 1.386 & 1.222 & 1.61 & 0.35 \\
1.00 & 1.0 & $10^{-7}$ & 8.29 & 9.65 & 21.3 & 3.87 & 9.55 & 1.387 & 1.221 & 1.63 & 0.36 \\
1.00 & 0.7 & $5 \times 10^{-8}$ & \multicolumn{9}{c}{Convection zone splits during simmering} \\
1.00 & 0.85 & $5 \times 10^{-8}$ & \multicolumn{9}{c}{Convection zone splits during simmering} \\
1.00 & 1.0 & $5 \times 10^{-8}$ & \multicolumn{9}{c}{Convection zone splits during simmering} \\
2.79& 0.7 & $10^{-6}$ & \multicolumn{9}{c}{Off-center carbon ignition} \\
2.79& 0.85 & $10^{-6}$ & \multicolumn{9}{c}{Off-center carbon ignition} \\
2.79& 1.0 & $10^{-6}$ & \multicolumn{9}{c}{Off-center carbon ignition} \\
2.79 & 0.7 & $10^{-7}$ & \multicolumn{9}{c}{Convection zone splits during simmering} \\
2.79 & 0.85 & $10^{-7}$ & \multicolumn{9}{c}{Convection zone splits during simmering} \\
2.79 & 1.0 & $10^{-7}$ & \multicolumn{9}{c}{Convection zone splits during simmering} \\
2.79 & 0.7 & $5 \times 10^{-8}$ & \multicolumn{9}{c}{Convection zone splits during simmering} \\
2.79 & 0.85 & $5 \times 10^{-8}$ & \multicolumn{9}{c}{Convection zone splits during simmering} \\
2.79 & 1.0 & $5 \times 10^{-8}$ & \multicolumn{9}{c}{Convection zone splits during simmering} \\
\noalign{\smallskip}
\tableline			
\end{tabular}
}
\end{table}		
\clearpage

\begin{table}
\parbox{\textwidth}{{\caption{Results for the models with a cooling age of 10 Gyr.}\label{Table_10Gyr}}}
\centering
\noindent\makebox[\textwidth]{ 
\begin{tabular}{lccccccccccc}
\tableline\tableline
\noalign{\smallskip}
$Z$ & $M_{\rm{WD}}^{0}$ & $\dot{M}$ & $\log T\rm{_{c}^{\, sim}}$ & $\rm{\log \rho_{c}^{\, sim}}$ & $\rm{\Delta t^{sim}}$ & $\rm{t}$ & $\rm{\log \rho_{c}}$ & $M_{\rm{WD}}$ & $M_{\rm{conv}}$ & $\rm{10^{3} \eta_{c}}$ & $\rm{10^{3} (\eta_{c}-\eta_{c,0}})$ \\
\noalign{\smallskip}
$[Z_{\odot}]$ & $[M_{\odot}]$ &  $[M_{\odot}\,\rm{yr}^{-1}]$ & $[\rm{K}]$ & $[\rm{g\,cm^{-3}}]$ & [$\rm{kyr}]$ & $[\rm{Myr}]$ & $[\rm{g\,cm^{-3}}]$ & $[M_{\odot}]$ & $[M_{\odot}]$ & $\mathrm{}$ & $\mathrm{}$ \\
\noalign{\smallskip}
\tableline
\noalign{\smallskip}
0.01 & 0.7 & $10^{-6}$ & \multicolumn{9}{c}{Off-center carbon ignition} \\
0.01& 0.85 & $10^{-6}$ & \multicolumn{9}{c}{Off-center carbon ignition} \\
0.01& 1.0 & $10^{-6}$ & \multicolumn{9}{c}{Off-center carbon ignition} \\
0.01 & 0.7 & $10^{-7}$ & 8.36 & 9.50 & 34.2 & 6.84 & 9.41 & 1.384 & 1.247 & 0.33 & 0.32\\
0.01 & 0.85 & $10^{-7}$ & 8.35 & 9.51 & 31.2 & 5.34 & 9.41 & 1.384 & 1.249 & 0.33 & 0.32\\
0.01 & 1.0 & $10^{-7}$ & 8.31 & 9.73 & 3.79 & 3.91 & 9.60 & 1.393 & 1.211 & 0.42 & 0.40 \\
0.01 & 0.7 & $5 \times 10^{-8}$ & 8.32 & 9.56 & 52.6 & 13.7 & 9.46 & 1.387 & 1.255 & 0.34 & 0.32 \\
0.01 & 0.85 & $5 \times 10^{-8}$ & 8.33 & 9.56 & 51.6 & 10.7 & 9.46 & 1.387 & 1.259 & 0.33 & 0.32 \\
0.01 & 1.0 & $5 \times 10^{-8}$ & 8.31 & 9.60 & 40.2 & 7.76 & 9.49 & 1.388 & 1.252 & 0.33 & 0.32 \\
0.10& 0.7 & $10^{-6}$ & \multicolumn{9}{c}{Off-center carbon ignition} \\
0.10& 0.85 & $10^{-6}$ & \multicolumn{9}{c}{Off-center carbon ignition} \\
0.10& 1.0 & $10^{-6}$ & \multicolumn{9}{c}{Off-center carbon ignition} \\
0.10 & 0.7 & $10^{-7}$ & 8.33 & 9.53 & 31.8 & 6.83 & 9.44 & 1.386 & 1.242 & 0.45 & 0.32 \\
0.10 & 0.85 & $10^{-7}$ & 8.33 & 9.54 & 31.4 & 5.36 & 9.45 & 1.386 & 1.244 & 0.45 & 0.32 \\
0.10 & 1.0 & $10^{-7}$ & 8.29 & 9.72 & 5.10 & 3.92 & 9.60 & 1.392 & 1.207 & 0.54 & 0.41 \\
0.10 & 0.7 & $5 \times 10^{-8}$ & 8.30 & 9.59 & 50.8 & 13.8 & 9.49 & 1.388 & 1.248 & 0.46 & 0.33 \\
0.10 & 0.85 & $5 \times 10^{-8}$ & 8.30 & 9.59 & 52.9 & 10.8 & 9.49 & 1.388 & 1.250 & 0.45 & 0.32 \\
0.10 & 1.0 & $5 \times 10^{-8}$ & 8.30 & 9.60 & 50.6 & 7.70 & 9.50 & 1.388 & 1.252 & 0.46 & 0.33 \\
0.33& 0.7 & $10^{-6}$ & \multicolumn{9}{c}{Off-center carbon ignition} \\
0.33& 0.85 & $10^{-6}$ & \multicolumn{9}{c}{Off-center carbon ignition} \\
0.33& 1.0 & $10^{-6}$ & \multicolumn{9}{c}{Off-center carbon ignition} \\
0.33 & 0.7 & $10^{-7}$ & 8.33 & 9.55 & 29.6 & 6.86 & 9.46 & 1.386 & 1.238 & 0.73 & 0.31 \\
0.33 & 0.85 & $10^{-7}$ & 8.33 & 9.56 & 28.2 & 5.36 & 9.46 & 1.386 & 1.239 & 0.73 & 0.31 \\
0.33 & 1.0 & $10^{-7}$ & 8.28 & 9.73 & 6.12 & 3.92 & 9.60 & 1.392 & 1.206 & 0.84 & 0.41 \\
0.33 & 0.7 & $5 \times 10^{-8}$ & 8.29 & 9.61 & 51.4 & 13.8 & 9.50 & 1.388 & 1.247 & 0.75 & 0.32 \\
0.33 & 0.85 & $5 \times 10^{-8}$ & 8.29 & 9.61 & 55.6 & 10.7 & 9.51 & 1.388 & 1.247 & 0.75 & 0.33 \\
0.33 & 1.0 & $5 \times 10^{-8}$ & 8.27 & 9.63 & 47.7 & 7.77 & 9.53 & 1.389 & 1.246 & 0.76 & 0.33 \\
1.00& 0.7 & $10^{-6}$ & \multicolumn{9}{c}{Off-center carbon ignition} \\
1.00& 0.85 & $10^{-6}$ & \multicolumn{9}{c}{Off-center carbon ignition} \\
1.00& 1.0 & $10^{-6}$ & \multicolumn{9}{c}{Off-center carbon ignition} \\
1.00 & 0.7 & $10^{-7}$ & 8.29 & 9.62 & 28.2 & 6.86 & 9.52 & 1.387 & 1.223 & 1.62 & 0.35 \\
1.00 & 0.85 & $10^{-7}$ & 8.29 & 9.61 & 30.1 & 5.36 & 9.53 & 1.386 & 1.226 & 1.62 & 0.35 \\
1.00 & 1.0 & $10^{-7}$ & 8.21 & 9.72 & 18.4 & 3.89 & 9.62 & 1.390 & 1.200 & 1.72 & 0.45 \\
1.00 & 0.7 & $5 \times 10^{-8}$ & \multicolumn{9}{c}{Convection zone splits during simmering} \\
1.00 & 0.85 & $5 \times 10^{-8}$ & \multicolumn{9}{c}{Convection zone splits during simmering} \\
1.00 & 1.0 & $5 \times 10^{-8}$ & 8.25 & 9.67 & 54.0 & 7.76 & 9.57 & 1.388 & 1.235 & 1.64 & 0.38 \\
2.79& 0.7 & $10^{-6}$ & \multicolumn{9}{c}{Off-center carbon ignition} \\
2.79& 0.85 & $10^{-6}$ & \multicolumn{9}{c}{Off-center carbon ignition} \\
2.79& 1.0 & $10^{-6}$ & \multicolumn{9}{c}{Off-center carbon ignition} \\
2.79 & 0.7 & $10^{-7}$ & \multicolumn{9}{c}{Convection zone splits during simmering} \\
2.79 & 0.85 & $10^{-7}$ & \multicolumn{9}{c}{Convection zone splits during simmering} \\
2.79 & 1.0 & $10^{-7}$ & 8.16 & 9.72 & 21.8 & 3.83 & 9.62 & 1.383 & 1.200 & 4.03 & 0.51 \\
2.79 & 0.7 & $5 \times 10^{-8}$ & \multicolumn{9}{c}{Convection zone splits during simmering} \\
2.79 & 0.85 & $5 \times 10^{-8}$ & \multicolumn{9}{c}{Convection zone splits during simmering} \\
2.79 & 1.0 & $5 \times 10^{-8}$ & \multicolumn{9}{c}{Convection zone splits during simmering} \\
\noalign{\smallskip}
\tableline			
\end{tabular}
}
\end{table}		

\end{document}